\documentclass[twocolumn,showpacs,superscriptaddress]{revtex4}

\usepackage{graphicx}% Include figure files
\usepackage{dcolumn}% Align table columns on decimal point
\usepackage{bm}% bold math
\usepackage{hyperref}
\usepackage{booktabs}
\usepackage{xcolor}
\usepackage[utf8]{inputenc}% for bibtex

\begin{document}

%\title{Constructing initial data for binary spinning neutron stars}
\title{Constructing Binary Neutron Star Initial Data
with \\ High Spins, High Compactness, and High Mass-Ratios}

\author{Wolfgang Tichy}
\author{Alireza Rashti}
\affiliation{Department of Physics, Florida Atlantic University,
             Boca Raton, FL 33431, USA}
\author{Tim Dietrich}
\affiliation{Nikhef, Science Park 105,
             1098 XG Amsterdam, The Netherlands}
\author{Reetika Dudi}
\affiliation{Max Planck Institute for Gravitational Physics (Albert Einstein Institute), 
             Am M\"uhlenberg 1, Potsdam 14476, Germany}
\affiliation{Theoretical Physics Institute, University of Jena,
             07743 Jena, Germany}
\author{Bernd Br\"ugmann}
\affiliation{Theoretical Physics Institute, University of Jena,
             07743 Jena, Germany}

%\date{$$Id: DNSdata1.tex,v 1.15 2018/12/12 14:06:36 wolf Exp $$}


% Sometimes we want to include preprint numbers, let's put them here

%\preprint{???}

%-------------------------------------------------------------------------
%Useful Definitions
%------------------------------------------------------------------------
%
\newcommand\be{\begin{equation}}
\newcommand\ba{\begin{eqnarray}}

\newcommand\ee{\end{equation}}
\newcommand\ea{\end{eqnarray}}
\newcommand\p{{\partial}}
\newcommand\remove{{{\bf{THIS FIG. OR EQS. COULD BE REMOVED}}}}
%

%-------------------------------------------------------------------------
\begin{abstract}
%-----------------------------------------------------------------------

The construction of accurate and consistent initial data for various binary
parameters is a critical ingredient for numerical relativity simulations of
the compact binary coalescence. In this article, we present an upgrade of
the pseudospectral SGRID code, which enables us to access even larger regions 
of the binary neutron star parameter space.
As a proof of principle, we
present a selected set of first simulations based on initial configurations
computed with the new code version. In particular, we simulate two millisecond
pulsars close to their breakup spin, highly compact neutron stars with
masses at about $98\%$ of the maximum supported mass of the employed equation 
of state, and an unequal mass systems with mass ratios even outside the range 
predicted by population synthesis models ($q = 2.03$). The discussed code extension
will help us to simulate previously unexplored binary configurations. This is
a necessary step to construct and test new gravitational wave approximants
and to interpret upcoming binary neutron star merger observations.
When we construct initial data, one has to specify various parameters, such as
a rotation parameter for each star. Some of these parameters do not have
direct physical meaning, which makes comparisons with other methods or models
difficult. To facilitate this, we introduce simple estimates for the initial
spin, momentum, mass, and center of mass of each individual star.

%-----------------------------------------------------------------------
\end{abstract}
%-----------------------------------------------------------------------

\maketitle

%%%%%%%%%%%%%%%%%%%%%%%%%%%%%%%%%%%%%%%%%%%%%%%%%%%%%%%%%%%
\section{Introduction}
%%%%%%%%%%%%%%%%%%%%%%%%%%%%%%%%%%%%%%%%%%%%%%%%%%%%%%%%%%%

In August 2017, the combined detection of a gravitational-wave (GW) signal
and the detection of electromagnetic (EM) signals across the whole spectrum
emitted from the same astrophysical source, a binary neutron star (BNS) merger,
initiated a new era of multi-messenger astronomy~\cite{TheLIGOScientific:2017qsa,GBM:2017lvd}.

While there are analytical models to describe the BNS coalescence as long as
the two stars are well separated, the highly non-linear regime around the
moment of merger is only accessible with full numerical relativity (NR)
simulations. These simulations allow us to study the dynamics, GW signal, and
possible EM counterparts, and are therefore required for a true
multi-messenger interpretation.

Most NR simulations are based on a 3+1-decomposition in which the
4-dimensional spacetime is foliated by spacelike hypersurfaces. This means
that for a successful numerical simulation one has to solve the Einstein
equations and the equations governing general relativistic matter on a
spacelike hypersurface as an initial condition; see e.g.,~\cite{Cook00a}
or~\cite{Tichy:2016vmv} and references therein. Generally, these initial
data have to provide configurations in which the stars are sufficiently
far away from each other to allow a study of the emitted GW signal, but
one also wants a distance short enough to avoid the computational cost
of too many orbits. Current
state-of-the-art BNS simulations reach from a few orbits up to 22 orbits
prior to merger~\cite{Haas:2016cop}. 

Given the diversity of the BNS population, one has to be able to construct
accurate initial data for a variety of different binary parameters for an
accurate interpretation of future detections. As an example, even relatively
small spins can, if neglected, lead to biases in the estimation of the
source properties,
e.g.,~\cite{Favata:2013rwa,Agathos:2015uaa,Samajdar:2019ulq}. This fact
together with the observation of a number of highly-spinning
neutron stars (NS), e.g., PSR J1748$-$2446ad~\cite{Hessels:2006ze} (the
fastest spinning NS, $716$~Hz), PSR J1807-2500B~\cite{Lynch:2011aa} (the
fastest spinning NS in a binary, $239$~Hz), and PSR
J1946+2052~\cite{Stovall:2018ouw} (the fastest spinning NS in a BNS system,
$59$~Hz), make the accurate modelling of spin effects indispensable.

Similarly, the observation of massive NSs $m_{\rm NS} > 2 M_\odot$,
e.g., PSR J0740+6620~\citep{Cromartie:2019kug} with
$m = 2.14^{+0.10}_{-0.09}$, shows that it is important to simulate
stars with high mass and thus high compactness.
Collisions of such massive stars might be almost indistinguishable
from the merger of small black holes (BH), since the amount of the ejected material
and consequently the brightness of the kilonova typically decrease for
high compactnesses and larger total masses~\cite{Dietrich:2016fpt}.
Additional simulations are needed to further
improve estimates of the prompt collapse 
threshold~\cite{Bauswein:2013jpa,Koppel:2019pys}, i.e.,
the mass at which the colliding neutron stars immediately form
a black hole. Such threshold mass estimates will
become particularly important once the increasing number of GW triggers
will no longer allow expensive EM follow-up campaigns for all potential
GW candidates and thus observational overhead needs to be reduced.

Finally, as shown in, e.g.,~\cite{Hinderer:2009ca,Dietrich:2016lyp,
Dietrich:2016fpt,Lehner:2016lxy,Zappa:2017xba,Kiuchi:2019lls}
the mass-ratio of a BNS system affects the GW and EM signals, where
higher mass-ratio systems are typically less GW but more EM-bright.
Based on the distribution of isolated, observed NSs, mass ratios up to
$q_{\max} \simeq 2.3$ are allowed, contrary to population synthesis models
which predict maximal values of $q_{\max} \simeq 1.8 {-} 1.9$,
e.g.,~\cite{Dominik:2012kk,Dietrich:2015pxa}. To date,
observationally confirmed is only a maximum mass ratio of
$q\sim 1.3$~\cite{Martinez:2015mya,Lazarus:2016hfu}, however,
this small value might purely be a selection effect due to the limited
number of observed BNS systems with well constrained individual masses.

Over the years, the numerical relativity community has developed a number of
codes for computing BNS initial data in certain portions of the parameter space.
Some of the best known codes are: 
the open source spectral code LORENE~\cite{lorene_web}
with non-public extensions, e.g.,~\cite{Kyutoku:2014yba},
the Princeton group's multigrid solver~\cite{East:2012zn},
BAM's multigrid solver~\cite{Moldenhauer:2014yaa,Dietrich:2018bvi},
the COCAL code~\cite{Tsokaros:2012kp,Tsokaros:2015fea},
SpEC's spectral solver Spells~\cite{Foucart:2008qt,Tacik:2015tja},
and the spectral code
SGRID~\cite{Tichy:2006qn,Tichy:2009yr,Tichy:2009zr,Dietrich:2015pxa}.
Recent developments include \cite{Ruter:2017iph,Vincent:2019qpd}.

These codes have been employed for a variety of studies in different
corners of BNS parameter space~\footnote{We refer here solely to simulations
based on consistent initial data, where \textit{consistent} refers to
simultaneously solving the Einstein Equations and the equations of general
relativistic hydrodynamics.} such as
spinning BNSs~\cite{Bernuzzi:2013rza,Dietrich:2015pxa,Dietrich:2016lyp,
Dietrich:2018upm,Most:2019pac,Tsokaros:2019anx,East:2019lbk},
precessing BNSs~\cite{Dietrich:2015pxa,Tacik:2015tja,Dietrich:2017xqb},
eccentricity reduced BNSs~\cite{Foucart:2015gaa,Haas:2016cop,Kiuchi:2017pte,Dietrich:2017aum,
Dietrich:2018upm,Foucart:2018lhe,Dietrich:2019kaq,Kiuchi:2019kzt},
highly eccentric BNSs~\cite{Chaurasia:2018zhg},
high mass BNSs, e.g.,~\cite{Dietrich:2018phi,Radice:2018pdn,Koppel:2019pys,Kiuchi:2019lls},
and high-mass ratio systems~\cite{Dietrich:2015pxa,Dietrich:2016hky}.

Despite these advances there are a number of possible configurations which,
so far, have been out of reach for the NR community, e.g., configurations with
total masses above $M\sim 3.4M_\odot$ have, to our knowledge, not been
simulated before. Similarly highly spinning and precessing systems close to
the breakup, or high mass ratio systems for soft EOSs have been out of reach
for the numerical relativity community.
All of these configurations are not excluded by
population synthesis models, e.g.,~\cite{Dominik:2012kk},
and, therefore, should be studied.
Even more importantly, extreme corners of the parameter
space have to be covered properly to be capable to test the reliability of
waveform approximants in regions in which they are employed 
during the analysis of GW signals, see e.g.~\cite{TheLIGOScientific:2017qsa,
Abbott:2018wiz,Abbott:2018exr,Abbott:2018lct}.

Thus, to be prepared for future BNS mergers, we have upgraded our initial
data code SGRID to allow a computation of BNS systems for large spins,
compactnesses, and mass ratios. As a proof of principle, we present the
first dynamical simulation of a BNS merger of two neutron stars close to the
break-up spin, a simulation with the highest mass ratio ($q=2.03$)
considered in numerical relativity for a soft equation of state,
and a simulation with two stars which
have $98\%$ of the maximum allowed mass for the employed EOS. In addition,
all these simulations employ initial data which have been eccentricity
reduced, which is an important ingredient for the production of high-quality
data.

The article is structured as follows, Sec.~\ref{BNSequations} gives an
overview of the equations which we need to solve to obtain consistent
initial configurations, Sec.~\ref{num_method} summarizes the numerical
methods employed in the upgraded SGRID code.
In Sec.~\ref{results_ID} we present first results for particular initial data
and in Sec.~\ref{results_Evo} preliminary simulations to prove the robustness
of our new methods.
We conclude in Sec.~\ref{summary}.
In addition, we present an empirical relation between the NS spin and 
SGRID's input parameters, the employed procedure for the eccentricity reduction, 
and a comparison between the old and new SGRID code in the Appendix. 

Throughout the article, we use geometric units in which $G=c=1$, as well as
$M_{\odot}=1$.
Latin indices such as $i$ run from 1 to 3 and denote spatial indices,
while Greek indices such as $\mu$ run from 0 to 3 and denote spacetime
indices.

%%%%%%%%%%%%%%%%%%%%%%%%%%%%%%%%%%%%%%%%%%%%%%%%%%%%%%%%%
\section{Binary neutron stars with spin in quasi-equilibrium}
\label{BNSequations}
%%%%%%%%%%%%%%%%%%%%%%%%%%%%%%%%%%%%%%%%%%%%%%%%%%%%%%%%%

We start by briefly describing the equations governing
BNSs in arbitrary rotation states in General Relativity.
These equations were derived in~\cite{Tichy:2011gw,Tichy:2012rp} and
extended to the case of eccentric orbits
in~\cite{Moldenhauer:2014yaa,Dietrich:2015pxa}; see also~\cite{rueter_2019} 
for a possible generalization. We refer to the
review of~\cite{Tichy:2016vmv} for further references.

We base our method on the Arnowitt-Deser-Misner (ADM) decomposition of
Einstein's equations~\cite{Arnowitt62} and rewrite the 4-metric $g_{\mu\nu}$
in terms of the 3-metric $\gamma_{ij}$, the lapse $\alpha$, the shift
$\beta^i$, and the extrinsic curvature $K^{ij}$. The NS matter is assumed to
be a perfect fluid with stress-energy tensor
\begin{equation}
T^{\mu\nu} = [\rho_0(1+\epsilon) + P] u^{\mu} u^{\nu} + P g^{\mu\nu}.
\end{equation}
Here $\rho_0$ is the rest-mass density (which is proportional to the number
density of baryons), $P$ is the pressure, $\epsilon$ is the internal energy
density divided by $\rho_0$ and $u^{\mu}$ is the 4-velocity of the fluid.
We also introduce the specific enthalpy
\begin{equation}
\label{spec_enthalpy_def}
h = 1 + \epsilon + P/\rho_0 .
\end{equation}
This quantity is useful because if we assume a polytropic equation of state
\begin{equation}
\label{polytrop}
P = \kappa \rho_0^{1+1/n} ,
\end{equation}
we can express the rest-mass density, the pressure and the internal energy in
terms of it. The $n$ here is known as the polytropic index, and $\kappa$
is a constant.
In this paper we consider several different EOSs,
all approximated by piecewise polytropes following~\cite{Read:2008iy}.
Each piece is defined within a certain interval in $\rho_0$ and
has its own $n_i$ and $\kappa_i$ in this interval.
Within each polytrope piece we find
\begin{eqnarray}
\label{rhoPeps_h_poly}
\rho_0   &=& \kappa_i^{-n_i} \left(\frac{h-k_i}{n_i+1}\right)^{n_i} ,\nonumber\\
P        &=& \kappa_i^{-n_i} \left(\frac{h-k_i}{n_i+1}\right)^{n_i+1} ,\nonumber\\
\epsilon &=& \frac{n_i}{n_i+1}(h-1) + \frac{k_i-1}{n_i+1} .
\end{eqnarray}
The constants $n_i$, $\kappa_i$, and $k_i$ have to be chosen such that $P$ and
$\epsilon$ are continuous across the $\rho_0$ intervals. For the $\rho_0$
interval starting at $\rho_0=0$, which corresponds to the outermost layer of
the star, one obtains $k=0$.

We express the fluid 4-velocity $u^{\mu}$ in terms of the 3-velocity
\begin{equation}
\label{utilde-proj}
^{(3)}\!\tilde{u}^i = h \gamma^i_{\nu}u^{\nu} ,
\end{equation}
which in turn is split into an irrotational piece $D^i \phi$
and a rotational piece $w^i$
\begin{equation}
\label{utilde-split}
^{(3)}\!\tilde{u}^i = D^i \phi + w^i ,
\end{equation}
where $D_i$ is the derivative operator compatible with the 3-metric
$\gamma_{ij}$.

In order to simplify the problem and to obtain elliptic equations we make
several assumptions. The first is the existence of an approximate
symmetry vector $\xi^{\mu}$, such that
\begin{equation}
\pounds_{\xi} g_{\mu\nu} \approx 0 .
\end{equation}
We also assume similar equations for scalar matter quantities such as $h$.
For a spinning star, however, $\pounds_{\xi} u^{\mu}$ is non-zero.
Instead we assume that
\begin{equation}
\label{assumption1}
\gamma_i^{\nu} \pounds_{\xi} \left(\nabla_{\nu}\phi\right) \approx 0 ,
\end{equation}
so that the time derivative of the irrotational piece of the
fluid velocity vanishes in corotating coordinates.
We also assume that
\begin{equation}
\label{assumption2}
\gamma_i^{\nu} \pounds_{\frac{\nabla \phi}{h u^0}} w_{\nu}
\approx 0 ,
\end{equation}
and
\begin{equation}
\label{assumption3}
^{(3)}\!\pounds_{\frac{w}{h u^0}} w_i \approx 0 ,
\end{equation}
which describe the fact that the rotational piece of the fluid velocity
is constant along the world line of the star center.

These approximations together with the additional assumptions of maximal
slicing
\begin{equation}
\gamma_{ij} K^{ij} = 0
\end{equation}
and conformal flatness
\begin{equation}
\label{conflat}
\gamma_{ij} = \psi^4 \delta_{ij}
\end{equation}
yield the following coupled equations:
\begin{equation}
\label{ham}
\bar{D}^2 \psi
 + \frac{\psi^5}{32\alpha^2} (\bar{L}B)^{ij}(\bar{L}B)_{ij}
 +2\pi \psi^5 \rho  =  0 ,
\end{equation}
\begin{equation}
\label{mom}
\bar{D}_j (\bar{L}B)^{ij}
 -(\bar{L}B)^{ij} \bar{D}_j \ln(\alpha\psi^{-6})
 -16\pi\alpha\psi^4 j^i  =  0 ,
\end{equation}
\begin{equation}
\label{dt_K_zero}
\bar{D}^2 (\alpha\psi) - \alpha\psi
\left[\frac{7\psi^4}{32\alpha^2}(\bar{L}B)^{ij}(\bar{L}B)_{ij}
      +2\pi\psi^4 (\rho+2S) \right]  = 0 ,
\end{equation}
\begin{equation}
\label{continuity4}
D_i \left[ \frac{\rho_0 \alpha}{h}(D^i \phi + w^i)
          -\rho_0 \alpha u^0 (\beta^i + \xi^i) \right] = 0 ,
\end{equation}
and
\begin{equation}
\label{h_from_Euler}
h = \sqrt{L^2 - (D_i \phi + w_i)(D^i \phi + w^i)}.
\end{equation}
Here
$(\bar{L}B)^{ij} = \bar{D}^i B^j + \bar{D}^j B^i
- \frac{2}{3} \delta^{ij} \bar{D}_k B^k$,
$\bar{D}_i = \partial_i$, and we have introduced
\begin{equation}
% \beta^i = B^i + (\Omega\times r)^i
% B^i = \beta^i + \Omega \epsilon^{ij3} (x^j - x_{CM}^j) .
B^i = \beta^i + \xi^i + \Omega \epsilon^{ij3} (x^j - x_{CM}^j) ,
\end{equation}
\begin{eqnarray}
\label{fluid_matter}
\rho   &=& \alpha^2 [\rho_0(1+\epsilon) + P] u^0 u^0 - P, \nonumber \\
j^i    &=& \alpha[\rho_0(1+\epsilon) + P] u^0 u^0
           (u^i/u^0 + \beta^i), \nonumber \\
S^{ij} &=& [\rho_0(1+\epsilon) + P]u^0 u^0
           (u^i/u^0 + \beta^i) (u^j/u^0 + \beta^j) \nonumber \\
       & &  + P \gamma^{ij} ,
\end{eqnarray}
\begin{eqnarray}
\label{uzero}
u^0 &=& \frac{\sqrt{h^2 + (D_i \phi + w_i)(D^i \phi + w^i)}}{\alpha h},
\nonumber \\
L^2 &=& \frac{b + \sqrt{b^2 - 4\alpha^4 [(D_i \phi + w_i) w^i]^2}}{2\alpha^2},
\nonumber \\
b &=& [ (\xi^i+\beta^i)D_i \phi - C]^2 + 2\alpha^2 (D_i \phi + w_i) w^i ,
\end{eqnarray}
where we sum over repeated spatial indices, and where
$C$ is a constant of integration that, in general,
can have a different value inside each star.

In addition to the construction of BNS configurations with
arbitrary spin, we also want to vary the eccentricity of the systems.
Thus, we follow the methods which we have developed
in~\cite{Moldenhauer:2014yaa} (see also~\cite{Tichy:2016vmv}).
In this approach, the symmetry vector has the form
\begin{eqnarray}
\label{ellinspiralKV}
\xi_{1/2}^0 &=& 1, \nonumber\\
\xi_{1/2}^i &=& \Omega (-x^2, x^1-x_{c1/2}^1, 0) +
\frac{v_r}{r_{12}} (x^i-x_{CM}^i) ,
\end{eqnarray}
where $\Omega$ is the orbital angular velocity chosen to lie along
the $x^3$-direction and $v_r$ is the radial velocity that needs to be
negative for a true inspiral.
Here, $x_{CM}^i$ denotes the center of mass position of the system,
$r_{12}$ the distance between the two star centers, and
\be
\label{eq:x_circlecenters}
x^1_{c{1/2}}
= x^1_{CM} + e (x^1_{C*1/2} - x^1_{CM})
\ee
depends on the eccentricity parameter $e$ and the location of the two star
centers $x^1_{C*1/2}$. The specific form of Eq.~(\ref{ellinspiralKV})
is derived from the following two assumptions:
(i) $\xi^{\mu}$ is along the motion of the star center.
(ii) Without inspiral, each star center moves along a segment of an
elliptic orbit at apoapsis that can be approximated by its inscribed circle.
The eccentricity parameter $e$ that appears in $x^1_{c1/2}$ and the radial
velocity $v_r$ is freely adjustable to obtain any orbit we want.
Using this new symmetry vector $\vec{\xi}_{1/2}$, we can still solve
the initial data equations with the same methods as described before.
Most important, in order to obtain a true inspiral orbit with low eccentricity,
we can adjust both $e$ and $v_r$, while $\Omega$ can be adjusted by other
means such as the "force balance" method discussed below. Or we can set
$e=0$ and directly adjust $\Omega$ and $v_r$ as discussed in
Appendix~\ref{EccRed-Proc}.

The elliptic equations (\ref{ham}), (\ref{mom}), (\ref{dt_K_zero}),
and (\ref{continuity4}) above have to be solved incorporating
the boundary conditions
\begin{equation}
\label{psi_B_alpha_BCs}
\lim_{r\to\infty}\psi = 1, \ \ \
\lim_{r\to\infty}B^i = 0, \ \ \
\lim_{r\to\infty}\alpha\psi = 1
\end{equation}
at spatial infinity, and
\begin{equation}
\label{starBC}
%(\bar{D}^i \phi)\bar{D}_i \rho_0 + \psi^{-2}\bar{w}^i \bar{D}_i \rho_0
%= h u^0 \psi^4 (\beta^i + \xi^i) \bar{D}_i \rho_0
(D^i \phi)D_i \rho_0 + w^i D_i \rho_0
= h u^0 (\beta^i + \xi^i) D_i \rho_0
\end{equation}
at each star's surface.
While, in general, the rotational piece of the fluid velocity $w^i$ can
be chosen freely, we will use the form
\begin{equation}
\label{w-choice}
w^i = \epsilon^{ijk}\omega^j (x^k - x^k_{C*}) ,
\end{equation}
which as demonstrated in~\cite{Tichy:2012rp} results in almost rigidly
rotating fluid configurations with low expansion and shear.
The parameter $x^k_{C*}$ denotes the location of the star center and $\omega^j$ is an
arbitrarily chosen vector that determines the star spin.
Summation over the repeated indices $j$ and $k$ is implied.

Once the equations (\ref{ham}), (\ref{mom}), (\ref{dt_K_zero}),
(\ref{continuity4}) and (\ref{h_from_Euler}) are solved we know
$h$ (and thus the matter distribution) and the fluid 3-velocity
$^{(3)}\!\tilde{u}^i$ via Eq.~(\ref{utilde-split}).
The 3-metric is obtained from
Eq.~(\ref{conflat}) and the extrinsic curvature is given by
\begin{equation}
K^{ij} =  \frac{1}{2\psi^4 \alpha}(\bar{L}\beta)^{ij} .
\end{equation}

% WT: I think we do not need this:
%
% Notice that Eq.~(\ref{h_from_Euler}) can also be written as
% \begin{eqnarray}
% \label{lnh}
% \ln h &+&
% \frac{1}{2}\ln
% \left[
% \alpha^2 - \left(\beta^i+\xi^i+\frac{w^i}{hu^0}\right)
%            \left(\beta_i+\xi_i+\frac{w_i}{hu^0}\right)
% \right]
% \nonumber \\
% &=& -\ln\Gamma + \ln(-C),
% \end{eqnarray}
% where we have introduced
% \begin{equation}
% \Gamma =
% \frac{
% \alpha u^0
% \left[
% 1-\left(\beta^i+\xi^i+\frac{w^i}{hu^0}\right)\frac{D_i\phi}{\alpha^2 hu^0}
%         - \frac{w_i w^i}{(\alpha hu^0)^2}\right]
% }
% {
% \sqrt{ 1 -  \left(\beta^i+\xi^i+\frac{w^i}{hu^0}\right)
%             \left(\beta_i+\xi_i+\frac{w_i}{hu^0}\right)\frac{1}{\alpha^2} }
% } .
% \end{equation}

%%%%%%%%%%%%%%%%%%%%%%%%%%%%%%%%%%%%%%%%%%%%%%%%%%%%%%%%%
\section{Numerical method}
\label{num_method}
%%%%%%%%%%%%%%%%%%%%%%%%%%%%%%%%%%%%%%%%%%%%%%%%%%%%%%%%%

The elliptic equations (\ref{ham}), (\ref{mom}), (\ref{dt_K_zero}) and
(\ref{continuity4}) together with the algebraic equation
(\ref{h_from_Euler}) are the main equations that we have to solve in order
to construct initial data. We do so using the SGRID
program~\cite{Tichy:2006qn,Tichy:2009yr,Tichy:2009zr,Dietrich:2015pxa}, which uses
pseudospectral methods to accurately compute spatial derivatives. We will
solve the whole set of equations using an iterative procedure where we first
solve the elliptic equations for a given matter distribution $h$, then update
the matter using the algebraic equation (\ref{h_from_Euler}), and then go
back to the first step.

%%%%%%%%%%%%%%%%%%%%%%%%%%%%%%%%%%%%%%%%%%%%%%%%%%%%%
\subsection{Surface fitting coordinates}

The matter inside each star is smooth. However, at the surface (at $h=1$),
$\rho_0$, $P$, and $\epsilon$ are not differentiable. So if we want to take
full advantage of a spectral method, the star surfaces should be domain
boundaries. However, when we update the matter distribution given by $h$
within our iterative approach the stars change shape. Hence the domain
boundaries have to be adjusted as well. In order to address this problem we
cover space by multiple domains each described by their own coordinates. For the
star domains these coordinates depend on a freely specifiable function which
will allow us to adapt the domain boundaries to the star surface.
In the past we have done this by making use of coordinates $(A,B,\varphi)$
introduced by Ansorg~\cite{Ansorg:2006gd}, which can cover all of space
using only 6 computational domains.
Here the coordinates $A$ and $B$ both range from 0 to 1, and $\varphi$ is a
polar angle measured around the $x$-axis. The coordinate transformations
contain freely specifiable functions $\sigma_{\pm}(B,\varphi)$ that can be
chosen such that domain boundaries coincide with the star surfaces.
Unfortunately, the coordinate transformation from Ansorg coordinates
$(A,B,\varphi)$ to Cartesian like coordinates $(x,y,z)$ is so complicated
that its inverse cannot be written down analytically. This makes it very
hard to adjust the functions $\sigma_{\pm}$ so that domain boundaries
coincide with the star surfaces. Furthermore, the coordinate transformation
is also singular. When we solve elliptic equations with a Newton scheme we
have to solve a linear problem for each Newton step.
However, the condition number of the matrices describing this linear problem
are very high due the coordinate singularities mentioned before.
This can lead to numerical inaccuracies that are hard to deal with.

For these reasons we have modified SGRID so that we can now use
surface fitting cubed sphere coordinates $(\lambda,A,B)$
that have no singularities anywhere.
\begin{figure}
\includegraphics[scale=0.5,clip=true]{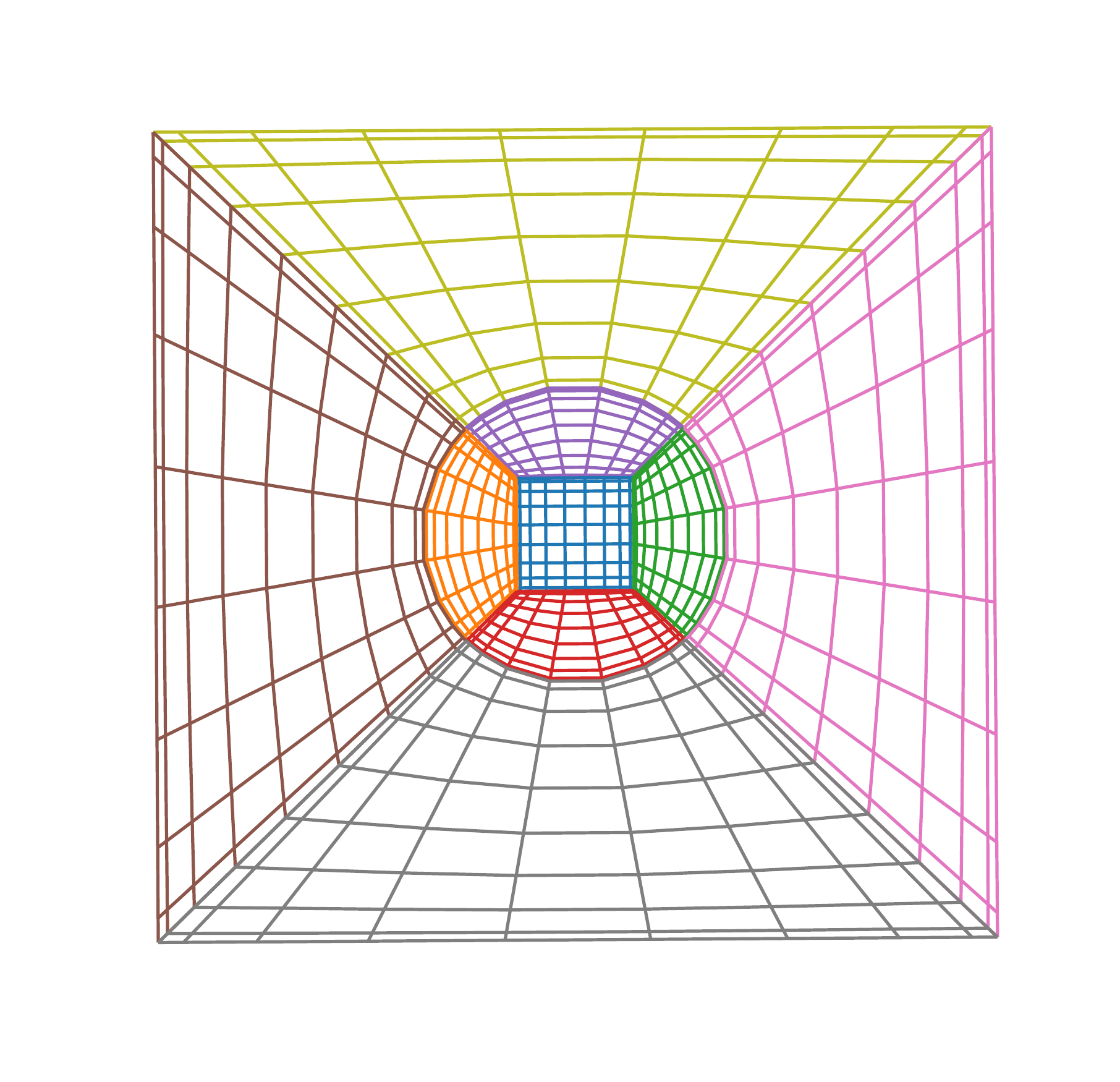}
\caption{\label{star1domains}
The plot shows some domains and their coordinate lines in the $xy$-plane.
Plotted are the domains inside and around neutron star 1. The star is
roughly spherical and covered by a central cube and six cubed sphere wedges,
four of which are shown because they intersect the $xy$-plane. The space
around the star is covered by six more domains to form a larger cube.
}
\end{figure}
In Fig.~\ref{star1domains} we show the coordinate lines in $z=0$ plane.
The star is covered by a central cube surrounded by several cubed sphere
wedges. The space around the star is covered by several more domains.
All domains together cover a larger cube containing the star and its
surroundings.
The coordinate transformation for the green wedge covering the star interior
to the right of the central cube is given by
\begin{eqnarray}
\label{CubedSphere_x}
x &=& x_{C*} +       (a_1-a_0) \lambda + a_0 ,           \nonumber \\
y &=& y_{C*} + \left[(a_1-a_0) \lambda + a_0\right] A  , \nonumber \\
z &=& z_{C*} + \left[(a_1-a_0) \lambda + a_0\right] B  ,
\end{eqnarray}
where
$\lambda \in [0,1]$, $A,B \in [-1,1]$ and
\begin{equation}
\label{outerCubedSphere+}
a_1 = \frac{\sigma_1(A,B)}{\sqrt{1+A^2+B^2}}, \ \ \
a_0 = const  .
\end{equation}
The function $\sigma_{1}(A,B)$ determines the shape of the star
surface. Notice that for $\sigma_{1}(A,B)=R_*=const$ we obtain a spherical
star surface with radius $R_*$. The coordinate lines in
Fig.~\ref{star1domains} are obtained for $B=0$.
The coordinate transformation for the other wedges inside the star
can be obtained by exchanging $x$ with $y$ or $z$ and
by possible sign changes of $a_1$ and $a_0$.
For example the red wedge covering the star interior below
the central cube is given by
\begin{eqnarray}
\label{CubedSphere_y}
y &=& y_{C*} +       (a_1-a_0) \lambda + a_0 ,           \nonumber \\
x &=& x_{C*} + \left[(a_1-a_0) \lambda + a_0\right] A  , \nonumber \\
z &=& z_{C*} + \left[(a_1-a_0) \lambda + a_0\right] B  ,
\end{eqnarray}
where now
\begin{equation}
\label{outerCubedSphere-}
a_1 = -\frac{\sigma_1(A,B)}{\sqrt{1+A^2+B^2}}, \ \ \
a_0 = -const  .
\end{equation}
The inverted wedges just outside the stars are obtained by reversing the
roles of $a_1$ and $a_0$. For the domain just below the red wedge we would have
\begin{equation}
\label{innerCubedSphere-}
a_0 = -\frac{\sigma_1(A,B)}{\sqrt{1+A^2+B^2}}, \ \ \
a_1 = -const ,
\end{equation}
while still using Eqs.~(\ref{CubedSphere_y}).
\begin{figure}
\includegraphics[scale=0.5,clip=true]{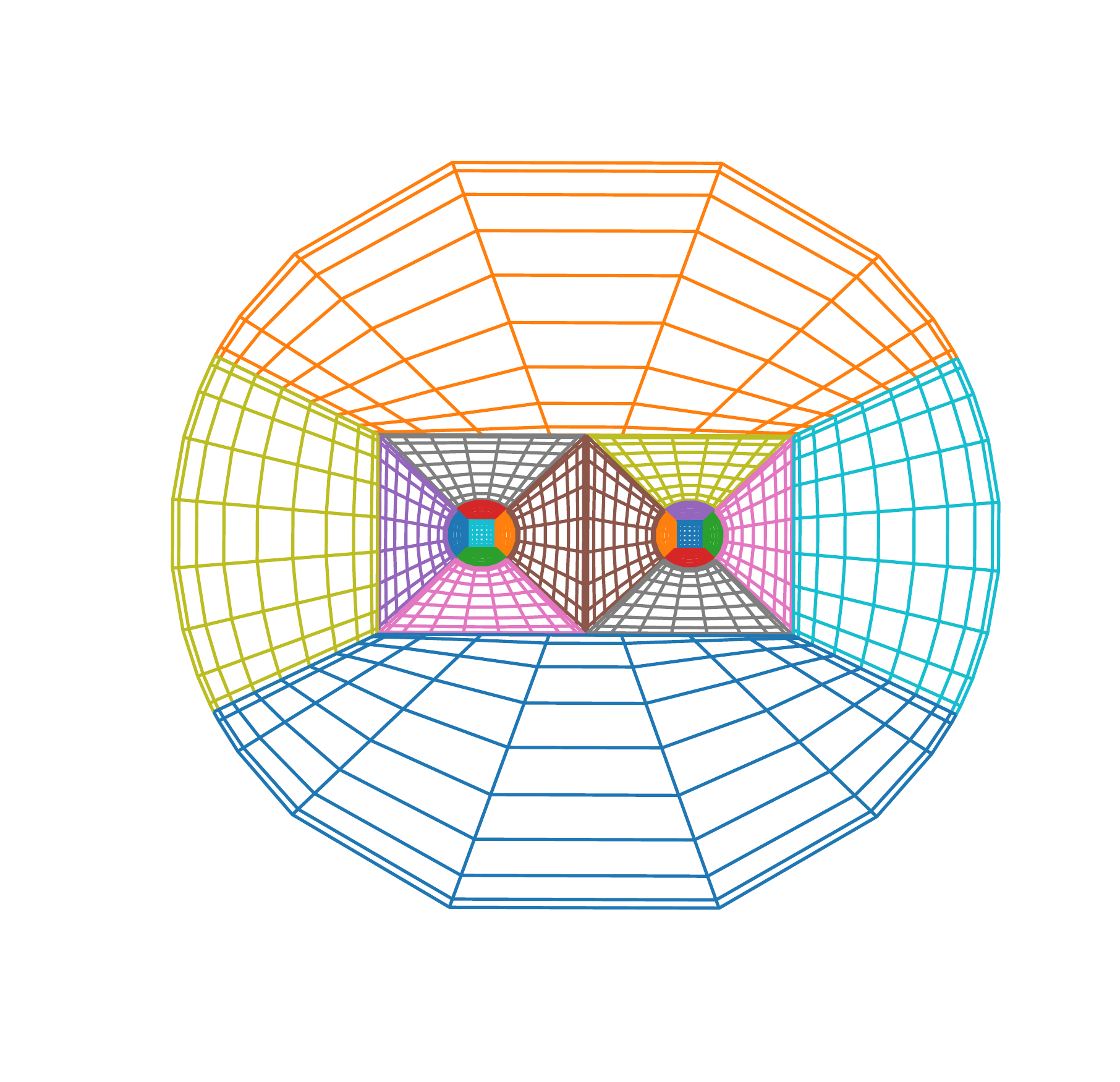}
\caption{\label{bothstars}
The plot shows the domains in and around both stars augmented by several
more domains. The result is a large sphere that covers both stars
and their surroundings.
}
\end{figure}

Fig.~\ref{bothstars} shows how two such larger cubes as in
Fig.~\ref{star1domains} can be put next to each other, and in turn be
surrounded by more wedges so as to cover a large sphere.
This sphere can in turn be surrounded by shells that can be obtained
by choosing
\begin{equation}
\label{CubedShell+}
a_0 = \frac{\sigma_{in}}{\sqrt{1+A^2+B^2}}, \ \ \
a_1 = \frac{\sigma_{out}}{\sqrt{1+A^2+B^2}} ,
\end{equation}
where $\sigma_{in}$ and $\sigma_{out}$ denote the inner and outer radius
of the shell. Since we have to impose the boundary conditions of
Eqs.~(\ref{psi_B_alpha_BCs}) at infinity
one should choose $\sigma_{out}$ to be very large.
For a given number of grid points in $\lambda$, this,
however, will result in poor resolution in the radial direction,
which could adversely affect the accuracy of our method.
For this reason we introduce yet another coordinate transformation.
If we define $r := \sqrt{x^2 + y^2 + z^2}$
and $L := \sigma_{out}-\sigma_{in}$, then Eqs.~(\ref{CubedSphere_x})
and (\ref{CubedShell+}) result in
\begin{equation}
r = L \lambda + \sigma_{in}.
\end{equation}
So if we want a domain that extends to a large radius it is
advantageous to replace the $\lambda$ coordinate with
\begin{equation}
\label{rho_of_lam}
\rho = \frac{\sigma_{out}}{L} \left( 1 - \frac{\sigma_{in}}{r} \right)
     = \frac{\sigma_{out}}{L}
       \left( 1 - \frac{\sigma_{in}}{L\lambda + \sigma_{in}} \right) .
\end{equation}
Then a quantity $\Psi$ that behaves as $\Psi \sim b_1 + b_2/r$
for large $r$, becomes $\Psi \sim c_1 + c_2\rho $, when expressed
in terms of $\rho$ (here $b_1$, $b_2$, $c_1$, $c_2$ are constants).
Thus, if within our spectral method we expand $\Psi(\rho)$ in terms of
Chebychev polynomials only the first few coefficients will be
non-negligible, which leads to a very good approximation when we keep only a
finite number of terms. This would not be the case if we used $\lambda$
as our coordinate since then $\Psi \sim d_1 + d_2/\lambda$, which is not a
polynomial in $\lambda$.

%%%%%%%%%%%%%%%%%%%%%%%%%%%%%%%%%%%%%%%%%%%%%%%%%%%%%
\subsection{Non-linear equations we have to solve}

In order to construct initial data we have to solve the elliptic equations
(\ref{ham}), (\ref{mom}), (\ref{dt_K_zero}), and (\ref{continuity4}). This is
done using SGRID's pseudospectral method as
in~\cite{Tichy:2006qn,Tichy:2009yr,Tichy:2009zr} where we use Chebychev
expansions and introduce grid points at the Chebychev extrema. Once the
number of grid points is chosen all derivatives are approximated by certain
linear combinations of the field values at the grid points. Such a
pseudospectral method is similar in spirit to finite differences but it uses
all grid points in one direction to approximate a derivative in this
direction and is, thus, much more accurate for smooth fields. Once all
derivatives have been discretized in this way, we end up with a set of
non-linear equations for all fields at all grid points. This system of
equations has the form
\begin{equation}
F_m (U) = 0,
\end{equation}
where the solution
vector $U$ is comprised of all the fields at all grid points, i.e.,
\begin{equation}
\label{full_discrete_eqns}
U=(\psi_0, \psi_1, ..., B^i_0, B^i_1, ...,
   (\alpha\Psi)_0, (\alpha\Psi)_1, ...,
   \phi_0, \phi_1, ...) ,
\end{equation}
where the subscripts label the grid points.
Note, however, that we also have to solve the algebraic equation
(\ref{h_from_Euler}), which is done in an iterative manner. We update
$h$ and thus the matter distribution after the elliptic equations
have been solved, and then the elliptic equations are solved again until
we reach a certain tolerance.
Because we have to iterate anyway, we do not solve the full system
of equations~(\ref{full_discrete_eqns}), but rather solve the
equations for $\psi$, $B^i$, $\alpha\Psi$, and $\phi$
individually one after the other within the overall iteration. Then the
non-linear system of equations we solve at once  is
\begin{equation}
\label{one_discrete_eqn}
f_m (u) = 0,
\end{equation}
where $u$ is now one of the six fields $\psi$, $B^i$, $\alpha\Psi$
or $\phi$.
To find the solutions we use a Newton-Raphson scheme where $u$
is updated according to
$u_{new} = u_{old} + x$ until a desired tolerance  has been reached.
As in any Newton scheme the correction $x$ is obtained by solving the
linearized equations
\begin{equation}
\label{linearEqs}
\frac{\partial f_m(u)}{\partial u^n } x^n = - f_m (u) .
\end{equation}
The challenging part of the method is then to find an efficient way
to solve this system of coupled linear equations. In the past,
Refs.~\cite{Tichy:2006qn,Tichy:2009yr,Tichy:2009zr,Dietrich:2015pxa}, when
using only 6 domains we were able to use a direct solver for the sparse
matrix $\frac{\partial f_m(u)}{\partial u^n }$. However now that
we are using 38 domains this is now longer efficient. We thus use an
iterative generalized minimal residual (GMRES) solver. 
This solver needs a good preconditioner, otherwise
it will take too many iterations to find a solution to the linearized
equations. A preconditioner is essentially an approximate inverse of the
matrix $\frac{\partial f_m(u)}{\partial u^n }$ that can be computed
efficiently. Here we use a block Jacobi
method~\cite{Reifenberger2013}, i.e., we keep only certain
blocks of the matrix $\frac{\partial f_m(u)}{\partial u^n }$ along the
diagonal. Such a block diagonal matrix $P$ is much easier to invert
and thus $P^{-1}$ can be used as a preconditioner. We obtain these blocks
by first dropping all entries in $\frac{\partial f_m(u)}{\partial u^n }$
that couple different computational domains. This results in 38
smaller blocks, each of which can be inverted more easily than the full
matrix $\frac{\partial f_m(u)}{\partial u^n }$. To further speed up
the computation of the preconditioner, we subdivide each box along both
the $A$ and $B$ coordinate directions so that we end up with
$2\times 2\times 38 = 152$ even smaller blocks along the diagonal of $P$,
which can now be readily inverted by a direct solver for
sparse matrices~\cite{Davis-Duff-1997-UMFPACK,Davis-Duff_UMFPACK_1999,
Davis_UMFPACK_V4.3_2004,Davis_UMFPACK_2004,umfpack_web}.
This block diagonal inverse $P^{-1}$ is used as our preconditioner for the
GMRES method, which allows us to solve the linear system in
Eq.~(\ref{linearEqs}), so that we can take a Newton step.

Since we solve Eqs.~(\ref{ham}), (\ref{mom}), (\ref{dt_K_zero}),
and (\ref{continuity4}) on 38 computational domains
we need interdomain boundary conditions that connect them.
In principle, these interdomain boundary conditions are very simple.
One imposes that each field and its normal derivative are continuous
across every interdomain boundary. These conditions are imposed by replacing
the elliptic equation at each boundary point by either
\begin{equation}
\label{IBC1}
u = u_{\mbox{adj}}
\end{equation}
or
\begin{equation}
\label{IBC2}
n^i \partial_i u = n^i \partial_i u_{\mbox{adj}} ,
\end{equation}
where $u_{\mbox{adj}}$ is the field value in the adjacent domain, and $n^i$
is the vector normal to the boundary.
Since both conditions have to be satisfied, one of them is imposed on
the boundary points on one side of the boundary and the other is imposed on
the other side in the adjacent domain. For the full system in
Eq.~(\ref{one_discrete_eqn}) it does not matter which condition is used on
which side. However, the preconditioner which contains blocks that come
from only one domain is sensitive to this issue. It turns out that if one
imposes condition~(\ref{IBC2}) on all sides of a domain, the block
corresponding to this domain has a determinant of zero
and thus cannot be inverted. In SGRID this problem is avoided by
making sure that condition~(\ref{IBC1}) is imposed on at least one boundary
of each domain. SGRID now has a facility that automatically finds
interdomain boundaries and imposes consistent conditions on them.

%%%%%%%%%%%%%%%%%%%%%%%%%%%%%%%%%%%%%%%%%%%%%%%%%%%%%
\subsection{Modification to conformal factor equation}
\label{ConFacMod}

The conformal factor $\psi$ has to satisfy Eq.~(\ref{ham}). Unfortunately
this equation is not guaranteed to have unique solutions. When this happens
the linear solver fails and one cannot find initial data. We have observed
that this does indeed happen when we try to construct initial data for very
compact stars. The problem can easily be
seen for zero shift ($B^i=0$) where Eq.~(\ref{ham}) takes the simple form
\begin{equation}
\label{hamB0}
\bar{D}^2 \psi = -2\pi\rho \psi^5 .
\end{equation}
If we linearize it we obtain
\begin{equation}
\label{lhamB0}
\bar{D}^2 \delta\psi = -10\pi\rho \psi^4 \delta\psi ,
\end{equation}
where $\delta\psi$ is the linearized conformal factor.
Linear elliptic equations of this type
are well known, and one can prove uniqueness only if the coefficient in
front of $\delta\psi$ on the right hand side is positive (see
e.g.~\cite{Gourgoulhon:2007ue}). However, since
both $\rho$ and $\psi$ are positive this coefficient is negative.
One can fix this problem by introducing a rescaled density
\begin{equation}
\label{rhobar}
\bar{\rho} = \psi^8 \rho
\end{equation}
so that Eq.~(\ref{ham}) becomes
\begin{equation}
\label{hamMOD}
\bar{D}^2 \psi = -2\pi \bar{\rho} \psi^{-3}
                 -\frac{\psi^5}{32\alpha^2} (\bar{L}B)^{ij}(\bar{L}B)_{ij} .
\end{equation}
If we keep $\bar{\rho}$ constant while we solve this equation, its
linearized version is
\begin{equation}
\label{lhamMOD}
\bar{D}^2 \delta\psi = +6\pi\bar{\rho} \psi^{-4} \delta\psi
-\frac{5\psi^4}{32\alpha^2} (\bar{L}B)^{ij}(\bar{L}B)_{ij} \delta\psi ,
\end{equation}
which now is guaranteed to have unique solutions for $B^i=0$.
The downside of this approach is that instead of solving the equation once,
one has to solve it iteratively. After each elliptic solve for $\psi$
one has to recompute $\bar{\rho}$ using Eq.~(\ref{rhobar}), and then solve
again until the changes in $\psi$ fall below a specified tolerance.
However, as described below we have to solve our system of equations
using an iterative approach anyway. We thus rescale $\rho$ according
to Eq.~(\ref{rhobar}) and only update $\bar{\rho}$ at the start of each
overall iteration.

%%%%%%%%%%%%%%%%%%%%%%%%%%%%%%%%%%%%%%%%%%%%%%%%%%%%%
\subsection{Modification to velocity potential equation near the star surface}

Notice that the elliptic equation (\ref{continuity4}) for the velocity
potential $\phi$ reduces to a first order equation at the star surface where
$\rho_0 \to 0$. In fact it reduces to Eq.~(\ref{starBC}) which we use as
boundary condition on the star surface. Nevertheless, in the star interior
we solve Eq.~(\ref{continuity4}). For challenging cases with high spins or
high masses we find numerical problems close to the star surface arising
from this equation. In these cases the first derivatives of $\phi$ can
develop visible kinks just inside the star surface. These kinks tend to
destabilize the overall iteration so that we cannot readily compute
initial data. We have found that we can smooth out these kinks by
replacing Eq.~(\ref{continuity4}) with
\begin{eqnarray}
\label{continuityMOD}
%\frac{c(\rho_0)\alpha}{h} D_i D^i \phi +
\frac{c(\rho_0)\alpha}{h} \psi^{-4} \partial^2\phi
+2\frac{\rho_0\alpha}{h} \psi^{-5} (\partial_i\psi) (\partial_i\phi) &&
\nonumber \\
+ \left(D_i \frac{\rho_0 \alpha}{h}\right) \left(D^i \phi\right)
+ D_i \left[ \frac{\rho_0 \alpha}{h}  w^i
            -\rho_0 \alpha u^0 (\beta^i + \xi^i) \right] &=& 0. 
\nonumber \\
\end{eqnarray}
In the first term we have added the function
\begin{equation}
c(\rho_0) = \rho_0 + \epsilon \rho_{0c}
\left(\frac{\rho_{0c}-\rho_0}{\rho_{0c}}\right)^4 ,
\end{equation}
which depends on a small number $\epsilon$ and on $\rho_{0c}$ which we
choose equal to $\rho_0$ at the star center. For $\epsilon=0$ we recover
Eq.~(\ref{continuity4}). But for positive $\epsilon$ the principal part of
Eq.~(\ref{continuityMOD}) now never vanishes. With this modification we are
able to find solutions also in more challenging cases. Notice that
$c(\rho_0) = \rho_0$ at the star center and that $c(\rho_0)$ differs from
$\rho_0$ mostly near the star surface. Since at the star surface we impose
the boundary condition (\ref{starBC}) that is derived from the unmodified
Eq.~(\ref{continuity4}), the modifications to $\phi$ are small.

The neutron star surfaces always coincide with domain boundaries so that it
is straightforward to impose the boundary condition (\ref{starBC}) for $\phi$
at each star surface. Notice, however, that Eq.~(\ref{continuity4}) and its
boundary condition in Eq.~(\ref{starBC}) do not uniquely specify a solution
$\phi$. If $\phi$ solves both Eqs.~(\ref{continuity4}) and (\ref{starBC})
$\phi + \mbox{const}$ will be a solution as well.
In order to obtain a unique solution we demand that $\phi$ is zero at the
star center, i.e.\ $\phi(x^i_{C*})=0$. We impose this condition by adding
the term $\phi(x^i_{C*})$ to Eq.~(\ref{continuityMOD}) on all grid points in
the cubic domain covering the star center.

%%%%%%%%%%%%%%%%%%%%%%%%%%%%%%%%%%%%%%%%%%%%%%%%%%%%%
\subsection{Iteration scheme}
\label{it-scheme}

The elliptic equations (\ref{hamMOD}), (\ref{mom}) and (\ref{dt_K_zero})
need to be solved in all domains, while the matter
equations~(\ref{continuityMOD}) and (\ref{h_from_Euler}) are solved only
inside each star.
In order to solve the elliptic Eqs.~(\ref{hamMOD}), (\ref{mom}),
(\ref{dt_K_zero}), and (\ref{continuityMOD}) we need a fixed domain
decomposition. However, the location of the star surfaces (where $h=1$) is
not known a priori, but rather determined by Eq.~(\ref{h_from_Euler}). For
this reason we use the following iterative procedure:
\begin{enumerate}
\item
We first find an initial guess for $h$ within each star, in practice we
simply choose Tolman-Oppenheimer-Volkoff solutions (see e.g.\ Chap.~23
in~\cite{Misner73}) for each. For the irrotational velocity potential we
choose $\phi = \Omega (x_{C*}^1 - x_{CM}^1) x^2$, where $x_{C*}^1$ and
$x_{CM}^1$ are the center of the star and the center of mass. We choose the
initial orbital angular velocity according to post-Newtonian theory.

\item
If the residual of Eq.~(\ref{continuityMOD}) is larger than
10\% of the combined residuals of Eqs.~(\ref{hamMOD}), (\ref{mom}), and
(\ref{dt_K_zero}), we solve
Eq.~(\ref{continuityMOD}) for $\phi$. We then reset $\phi$ to
$\phi = 0.2 \phi_{ell} + 0.8 \phi_{old}$, where $\phi_{ell}$ is the just
obtained solution of Eq.~(\ref{continuityMOD}) and $\phi_{old}$ is the
previous value of $\phi$.

\item
Next we solve the 5 coupled elliptic equations
(\ref{hamMOD}), (\ref{mom}), and (\ref{dt_K_zero}) for
$\Psi_{ell} = (\psi, B^i, \alpha)_{ell}$.
We then set $\Psi=(\psi, B^i, \alpha)$ to
$\Psi = 0.2 \Psi_{ell} + 0.8 \Psi_{old}$.

\item\label{it-Omega}
In order to solve Eq.~(\ref{h_from_Euler}) we need
to know the values of the constants $C_{\pm}$ in each star
as well as $\Omega$ and $x_{CM}^1$.
We want to keep the star centers $x_{C*{1/2}}^1$ fixed at their
initial position, so that the stars do not drift around during the
iterations. The location of each star center is given by
$\partial_1 h|_{x_{C*{1/2}}^1} = 0$.
Note that this condition depends on $\Omega$ and $x_{CM}^1$.
One strategy to find $\Omega$ and $x_{CM}^1$ is thus to use a root finder
to adjust $\Omega$ and $x_{CM}^1$ until this condition is satisfied. This
method is known as "force balance". In some cases we use this
force balance method. However, often it is advantageous to fix
$\Omega$ by other means, e.g.\ by using an eccentricity reduction procedure
as described in Appendix~\ref{EccRed-Proc}.
In this case one only needs to find
$x_{CM}^1$. This can be achieved by adjusting $x_{CM}^1$ such that
the $y$-component of the ADM linear momentum is zero. Here the $y$-direction
denotes the direction perpendicular to both the orbital angular momentum and
the line connecting the two star centers.

\item
Next, we use Eq.~(\ref{h_from_Euler}) to update $h$ in each star, while at
the same time adjusting $C_{\pm}$ such that the rest mass of each star
remains constant. The domain boundaries need to be adjusted (by changing the
surface functions such as $\sigma_{1}(A,B)$ in Eq.~(\ref{outerCubedSphere+}))
so that they remain at the star surfaces, which change whenever $h$ is
updated.

\item
We then evaluate the residuals [i.e.\ the $L^2$-norm of the left
hand sides of Eqs.~(\ref{hamMOD}), (\ref{mom}), (\ref{dt_K_zero}), and
(\ref{continuityMOD})]. If the combined residual
is below a prescribed tolerance we are done and exit the iteration
at this point.

\item
In order to ensure that the star centers always remain at their original
position we use a root finder to find the locations where
$\partial_i h = 0$. We then translate $h$ (and all other matter variables
such as $\rho_0$ and $P$) by the amount necessary to bring them back to
the original $x_{C*{1/2}}^1$.

\item
Finally we go back to step 2.
\end{enumerate}

%%%%%%%%%%%%%%%%%%%%%%%%%%%%%%%%%%%%%%%%%%%%%%%%%%%%%
\section{Mass, center, momentum, and spin of individual stars}
\label{localQuantities}
%%%%%%%%%%%%%%%%%%%%%%%%%%%%%%%%%%%%%%%%%%%%%%%%%%%%%

In General Relativity no unambiguous definitions for the mass and spin of an
individual star in a binary system exist. Here we introduce easy
to compute estimates for such local quantities; 
see also e.g.~\cite{Campanelli:2006fy,Tacik:2015tja,Dietrich:2016lyp}.

A star mass estimate can be obtained from
\begin{equation}
\label{Mdef}
M := -\frac{1}{2\pi} \int_* f^{kl}\partial_k\partial_l\psi \sqrt{f} d^3 x .
\end{equation}
This equation has the same form as the ADM mass for conformally flat metrics,
however, the integration runs only over the star.
Here $f_{ij} = \delta_{ij}$ is the flat conformal metric. We find that this
quantity is much closer to the mass of an individual star with the same
baryonic mass than an analog definition using the physical metric
$\gamma_{ij} = \psi^4 f_{ij}$ in place of $f_{ij}$. Also, if one considers
the special case of the Schwarzschild metric in conformally flat isotropic
coordinates, the above definition yields the correct mass, while a
definition using the physical metric $\gamma_{ij}$ would give a mass that is
too large.

Since the above integral seems to capture the mass aspect of a star, we
introduce an analogous integral to define the center of the star
\begin{equation}
\label{Rdef}
R_c^i := -\frac{1}{2\pi}
          \int_* \frac{x^i-x_{CM}^i}{M} f^{kl}\partial_k\partial_l\psi
              \sqrt{f} d^3 x .
\end{equation}
This is essentially the same integral, except now weighted with the
coordinate $x^i$ divided by the mass $M$.

In order to obtain a momentum estimate we start with
\begin{equation}
P(k) = \frac{1}{8\pi} \oint_* K_{ij} k^i n^j \sqrt{q} d^2 y ,
\end{equation}
which is again inspired by the definitions for the ADM linear and angular
momenta (see e.g~\cite{Gourgoulhon:2007ue}). However
the integration here runs only over the surface of the star. Here,
$n^i$ is the normal vector of the star surface and $q$ is the determinant
of the metric induced on the surface by the physical metric and is given by
\begin{equation}
 q_{ij} = \gamma_{ij} - n_i n_j,\ \
  n_i = \gamma_{ij} n^j,\ \
  n_i n^i = 1 .
\end{equation}
The vector $k^i$ is a symmetry vector that could be a translational or
rotational Killing vector resulting in linear or angular momentum.
However since no exact Killing vectors will exist in the case of binaries,
and also to keep things simple, we will construct $k^i$ from the
coordinate unit vectors $(1,0,0)$, $(0,1,0)$, $(0,0,1)$ for the case of
linear momentum, and from the coordinate rotation vectors
$(1,0,0)\times\vec{r}$, $(0,1,0)\times\vec{r}$, $(0,0,1)\times\vec{r}$,
where $\vec{r}=(x,y,z)$.
For linear momentum and angular momentum about $\vec{x}_{CM}$ we thus obtain
\begin{equation}
\label{Pdef}
P^i := \frac{1}{8\pi} \oint_* K_{il} n^l \sqrt{q} d^2 y ,
\end{equation}
and
\begin{equation}
\label{Jdef}
J^i := \frac{1}{8\pi} \oint_* K_{kl} n^l
       \epsilon^{ijk} (x^j-x_{CM}^j) \sqrt{q} d^2 y .
\end{equation}
Notice that we would obtain the same results for $P^i$ and $J^i$ if we had
defined them using the conformal $\bar{K}_{ij} = \psi^2 K_{ij}$ while at the
same time defining $q_{ij}$ to be the metric induced by the conformal metric
$f_{ij}$. Also note that the usual surface integrals at infinity for ADM
linear momentum and angular momentum can be converted into volume integrals.
These volume integrals have support only within the stars, so that a natural
definition for the star momentum is just this volume integral over the
star. Furthermore each such volume integral over the star can be rewritten in
terms of a surface integral over the star surface. The expressions for the
resulting surface integrals are the same as Eqs.~(\ref{Pdef}) and
(\ref{Jdef}). This means that for a binary the $J^i$ for each star
will add up to the total ADM angular momentum.
These facts should give us a measure of confidence in the definitions
(\ref{Pdef}) and (\ref{Jdef}), probably more confidence than in the
mass definition (\ref{Mdef}), where such arguments do not apply.

Now that we can compute linear and angular momentum as well as the star center,
we can define the star spin in the usual way as
\begin{equation}
\label{Sdef}
S^i := J^i - \epsilon^{ijk} R_c^j P^k .
\end{equation}
The biggest uncertainty in this expression comes from $R_c^j$. However,
since $R_c^j$ computed using Eqs.~(\ref{Mdef}) and (\ref{Rdef}) is a ratio
of integrals, errors in the mass definition may at least partially divide
out.

%%%%%%%%%%%%%%%%%%%%%%%%%%%%%%%%%%%%%%%%%%%%%%%%%%%%%
\section{Numerical results: Initial data construction}
\label{results_ID}
%%%%%%%%%%%%%%%%%%%%%%%%%%%%%%%%%%%%%%%%%%%%%%%%%%%%%

%%%%%%%%%%%%%%%%%%%%%%%%%%%%%%%%%%%%%%%%%%%%%%%%%%%%%
\subsection{Initial data sequences}

\begin{figure}[ht]
\includegraphics[scale=0.5,clip=true]{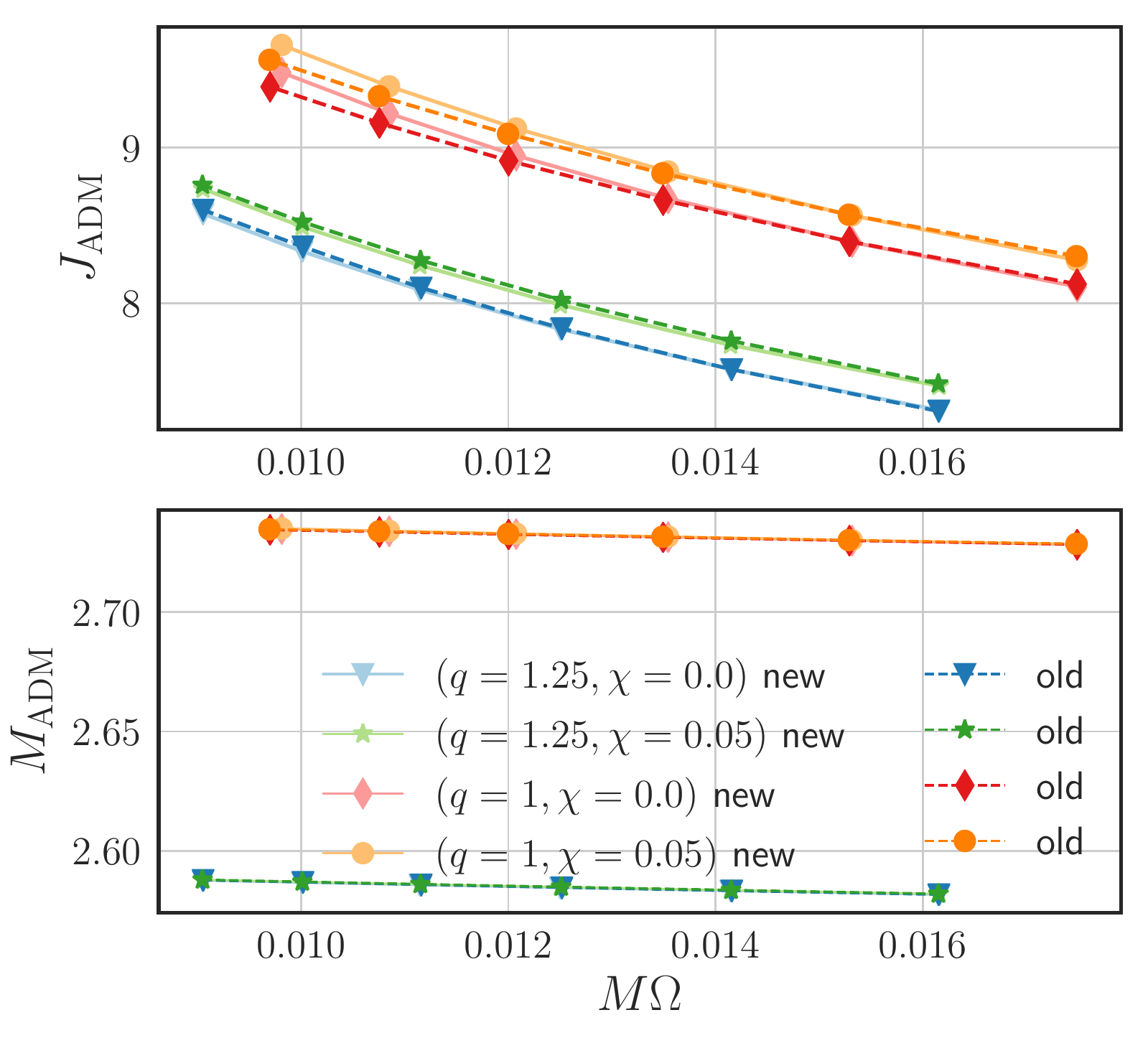}
\caption{\label{jadm} ADM-angular momentum ($J_{ADM}$) and mass ($M_{ADM}$)
as a function of the angular orbital velocity $M\Omega$.
Solid lines refer to results obtained with the new SGRID code,
dashed lines are obtained with the old implementation.
}
\end{figure}

As a first test of the upgraded SGRID code,
we compute for four sets of binary parameters initial data sequences
comparing the results of the old and the new SGRID implementation
(see also Appendix~\ref{old_vs_new_sgrid}).
All configurations employ a piecewise-polytropic fit of the
SLy EOS~\cite{Read:2008iy,Dietrich:2015pxa}. The gravitational masses are
either $m_1 = m_2 = 1.375 M_{\odot} $ with mass ratio $q = 1$, or
$m_1 = 1.445 M_{\odot},  m_2 = 1.156 M_{\odot} $ with mass ratio
$q = 1.25$, combined with the dimensionless spins $\chi=0$ and $\chi=0.05$.
Figure~\ref{jadm}
shows the ADM angular momenta ($J_{\rm ADM}$, top panel)
and ADM masses ($M_{\rm ADM}$, bottom panel) as a functions of
orbital velocity $(M\Omega)$ for all four configurations and for the new
and old SGRID code (dashed lines). 
The slight differences for large separations, i.e., 
small orbital frequencies, might be due to different eccentricities 
of the individual setups. 

\begin{figure}[ht]
\includegraphics[scale=0.48,clip=true]{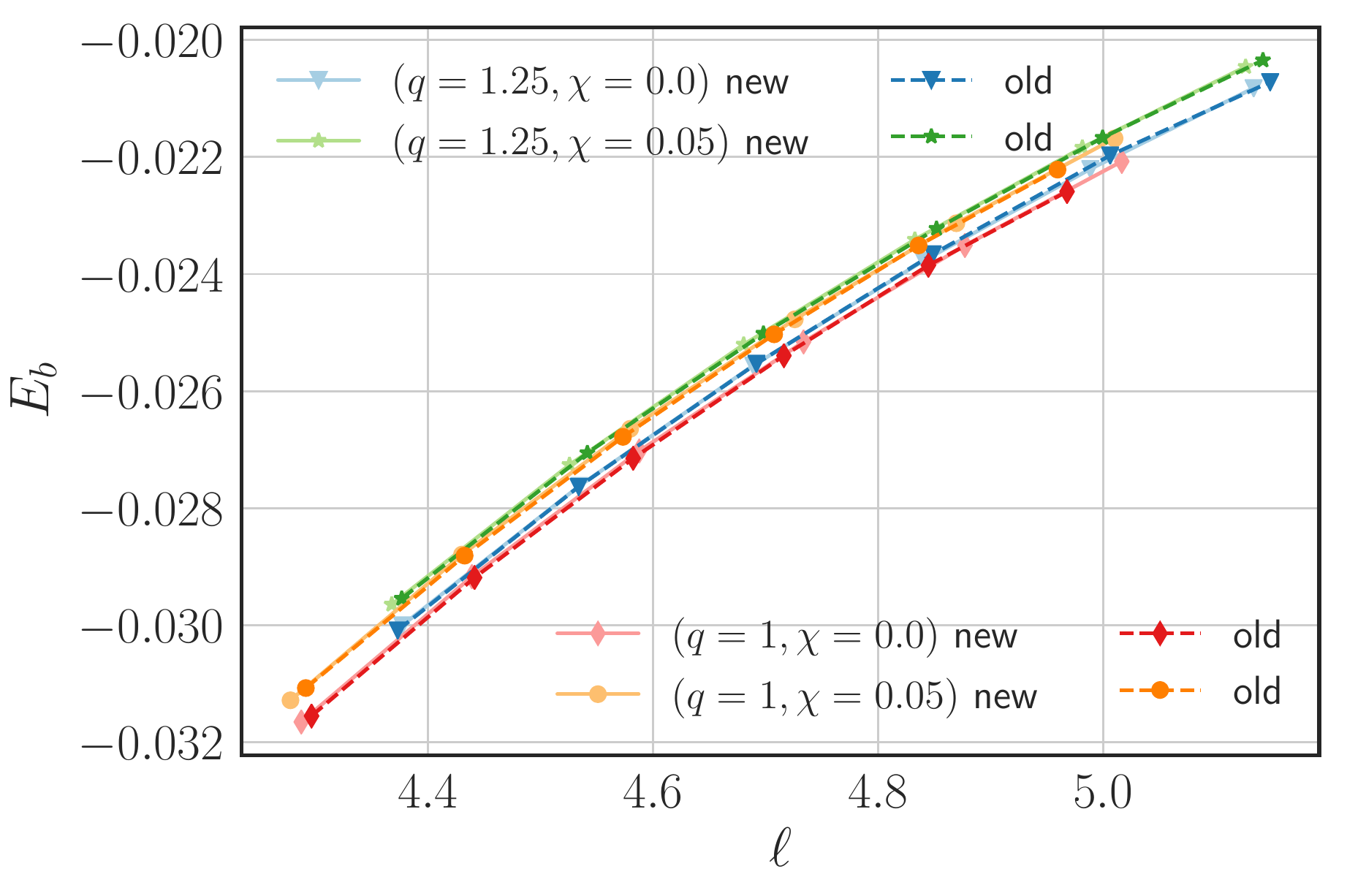}
\caption{\label{ebel}
Binding energy $E_b$ as function of the reduced orbital angular momentum $\ell$
for different configurations as discussed in the main text. Solid lines show
the results for the new, and dashed lines for the old SGRID implementation.
}
\end{figure}

In Fig.\ \ref{ebel}, we plot the binding energy
\begin{equation}
E_b = \frac{1}{\nu} \left(\frac{M_{ADM}}{M} -1\right)
\end{equation}
versus the reduced orbital angular momentum
\begin{equation}
l = \frac{L}{\nu M^2} = \frac{J_{ADM} - S_1 - S_2}{\nu M^2} .
\end{equation}
Here $\nu = m_1 m_2/M^2$ is the symmetric mass ratio, $M$ is
the total mass and $S_{1,2}$ are the individual spin magnitudes.

In Fig.\ \ref{ebel}, the solid lines represent the new SGRID data while
the dashed curves represent results obtained with the previous code version.
We find that both results are in good agreement with each other,
which validates our new implementation.

%%%%%%%%%%%%%%%%%%%%%%%%%%%%%%%%%%%%%%%%%%%%%%%%%%%%%
\subsection{Testing our spin definition for individual stars}

\begin{figure}[ht]
\includegraphics[scale=0.4,clip=true]{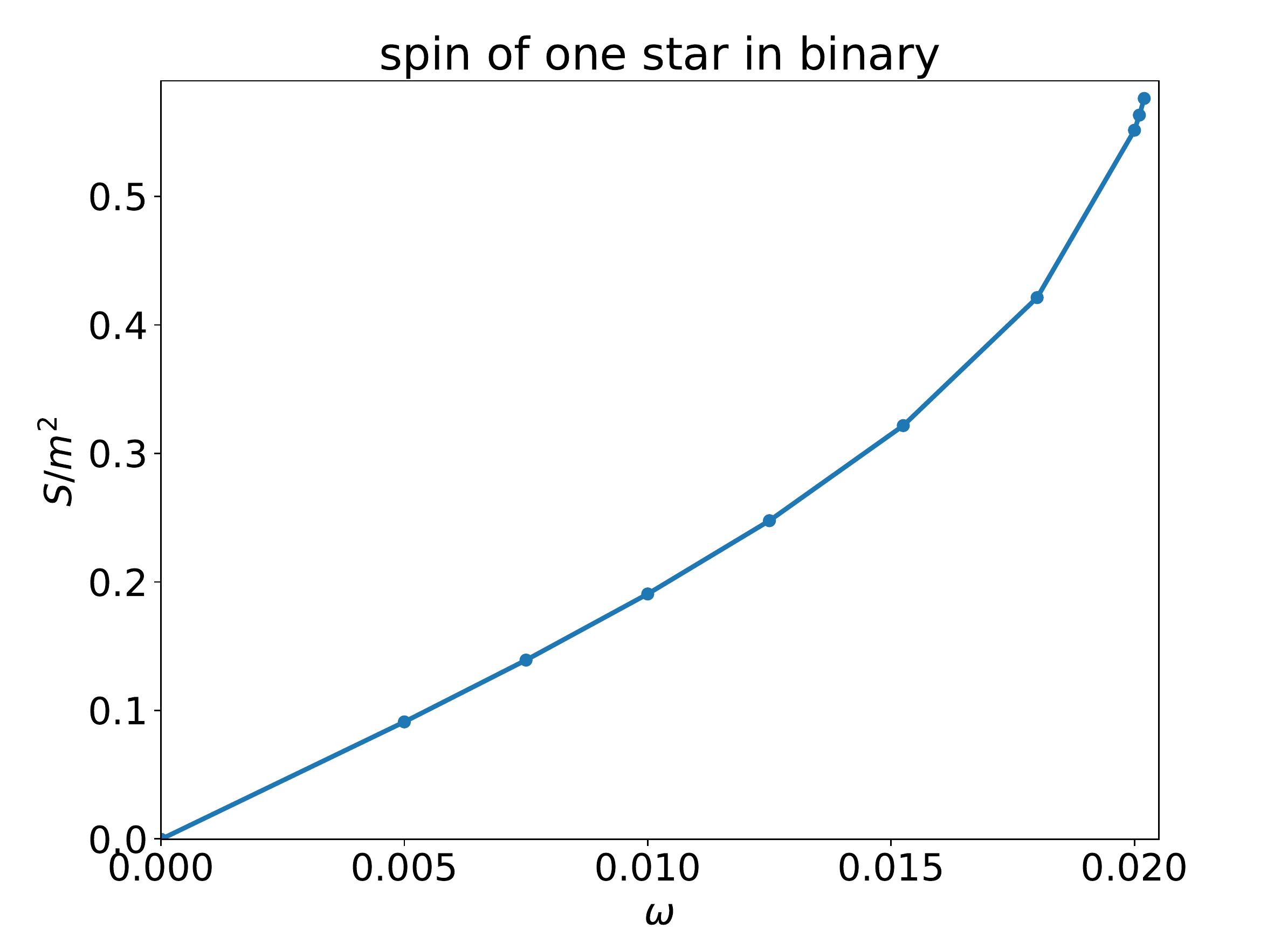}
\caption{\label{M_S_forPoly}
Dimensionless spin of one star in an equal mass binary with
$m_{1/2}=1.64$, $r_{12}=47.2$, using a polytropic equation
of state with $\kappa=123.6$, $n=1$.
}
\end{figure}

In Tab.~\ref{tab:M_S_forPoly} we show results of our mass and spin
definitions, Eqs.(\ref{Mdef}) and (\ref{Sdef}),
for the case of a single star and a BNS system with and without spin.
We see that the mass definition (\ref{Mdef}) for an individual star
differs from the ADM mass in isolation (which is $m=1.64$)
by about 1\% in the case of a binary, and is exact only for a single
non-spinning star.
The spin definition is exact for a single star and the spin estimates
for binaries are very likely better than 1\% 
accurate~\footnote{As we can see $S_1$ is not exactly zero for $\omega=0$
in the case of binaries. In Eq.~(F10) of~\cite{Marronetti:2003gk} it is
demonstrated that one may expect a non-zero spin angular velocity $\Omega_s$
even for irrotational stars. However, our initial data formulation as well
as our new spin definition differ from the approach
in~\cite{Marronetti:2003gk}. To compare we can estimate the moment of
inertia as $I\sim S_1/\omega \sim 50$ from the slope of
Fig.~\ref{M_S_forPoly}. So $S_1=-0.0007$ corresponds to
$\Omega_s=S_1/I=-1.5\times 10^{-5}$ and thus $\Omega_s/\Omega=-0.003$, which
is much smaller than the $0.06$ predicted by Eq.~(F10)
of~\cite{Marronetti:2003gk}.},
because of the partial cancellation of errors in $R^i_{c}$ discussed
in Sec.~\ref{localQuantities}.
\begin{table}
\begin{tabular}{l|ccc}
$m_0=1.7745$ $\rightarrow$ TOV $m=1.64$ & $M_1$ & $S_1$      & $S_1/m^2$\\
\hline
one non-spinning star ($\omega=0$)      & 1.640 & 0          & 0         \\
one spinning star ($\omega=0.01525$)    & 1.646 & $+0.8706$  & $+0.3237$ \\
two non-spinning stars ($\omega=0$)     & 1.620 & $-0.0007$  & $-0.0003$ \\
two spinning stars ($\omega=0.01525$)   & 1.626 & $+0.8652$  & $+0.3217$ \\
\end{tabular}
\caption{
Mass and spin estimates for the case of a polytropic equation of state
$P = \kappa \rho_0^{1+1/n}$ with $\kappa=123.6$, $n=1$.
For a binary with separation of $47.2$ (with $\Omega=0.005096$),
the mass definition differs from the ADM mass in isolation by about 1\%.
The spin definition is exact for a single star, and it is almost the same
for a single star and a star in a binary if the spin angular velocity
($\omega=0.01525$) is the same in both cases.
}
\label{tab:M_S_forPoly}
\end{table}

In Fig.~\ref{M_S_forPoly} we show the spin computed with Eq.~(\ref{Sdef})
versus the spin angular velocity $\omega$ for an equal mass binary
with equal spins aligned with the orbital angular momentum.
In this case we can reach a spin of $S/m^2 = 0.5763$
at $\omega=0.0202$, which is slightly
beyond the mass shedding limit of about $0.5705$ for a single star with this
polytropic equation of state~\cite{Ansorg:2003br}.
If we further increase $\omega$, SGRID fails. This happens because during the
iterations the star expands far into the domains that are supposed
to be outside of the star such that it is impossible to adjust our domains to be
surface fitting. We think that this is not a true failure of the the program
and should be expected to happen, since the stars will shed mass at these
spin angular velocities.

%%%%%%%%%%%%%%%%%%%%%%%%%%%%%%%%%%%%%%%%%%%%%%%%%%%%%
\section{Numerical results: Dynamical Evolutions}
\label{results_Evo}
%%%%%%%%%%%%%%%%%%%%%%%%%%%%%%%%%%%%%%%%%%%%%%%%%%%%%

%%%%%%%%%%%%%%%%%%%%%%%%%%%%%%%%%%%%%%%%%%%%%%%%%%%%%
\subsection{Evolving millisecond pulsars}

\begin{table*}
\begin{tabular}{l|cccc|cc}
resolution & $m_{1,2}^{\rm quasi-loc.}$ & $\chi_{1,2}^{\rm quasi-loc.}$ &
             $m_{1,2}^{\rm singl.\ star}$ & $\chi_{1,2}^{\rm signl.\ star}$ & 
             $M_{\rm ADM}$ & $J_{\rm ADM}$ \\
             \hline
$22\times 22 \times 22$ &1.346800 & 0.59466 & 1.364748  & 0.57536 & 2.711566 & 9.8464958 \\
$26\times 26 \times 26$ &1.346948 & 0.59474 & 1.365494  & 0.57504 & 2.711535 & 9.8494049\\
\end{tabular}
\caption{\label{tab:millisecond}
Mass and dimensionless spin for different resolutions for the binary
millisecond configuration, as well as the ADM mass and angular momentum. The
NS spin and mass estimates are computed from the quasi-local measures
introduced before and from a comparison to single star values estimated from
isolated stars with the same EOS, baryonic mass, and rotational velocity as
the individual constituents of the binary system.
}
\end{table*}

\begin{figure}[ht]
\includegraphics[width=\columnwidth]{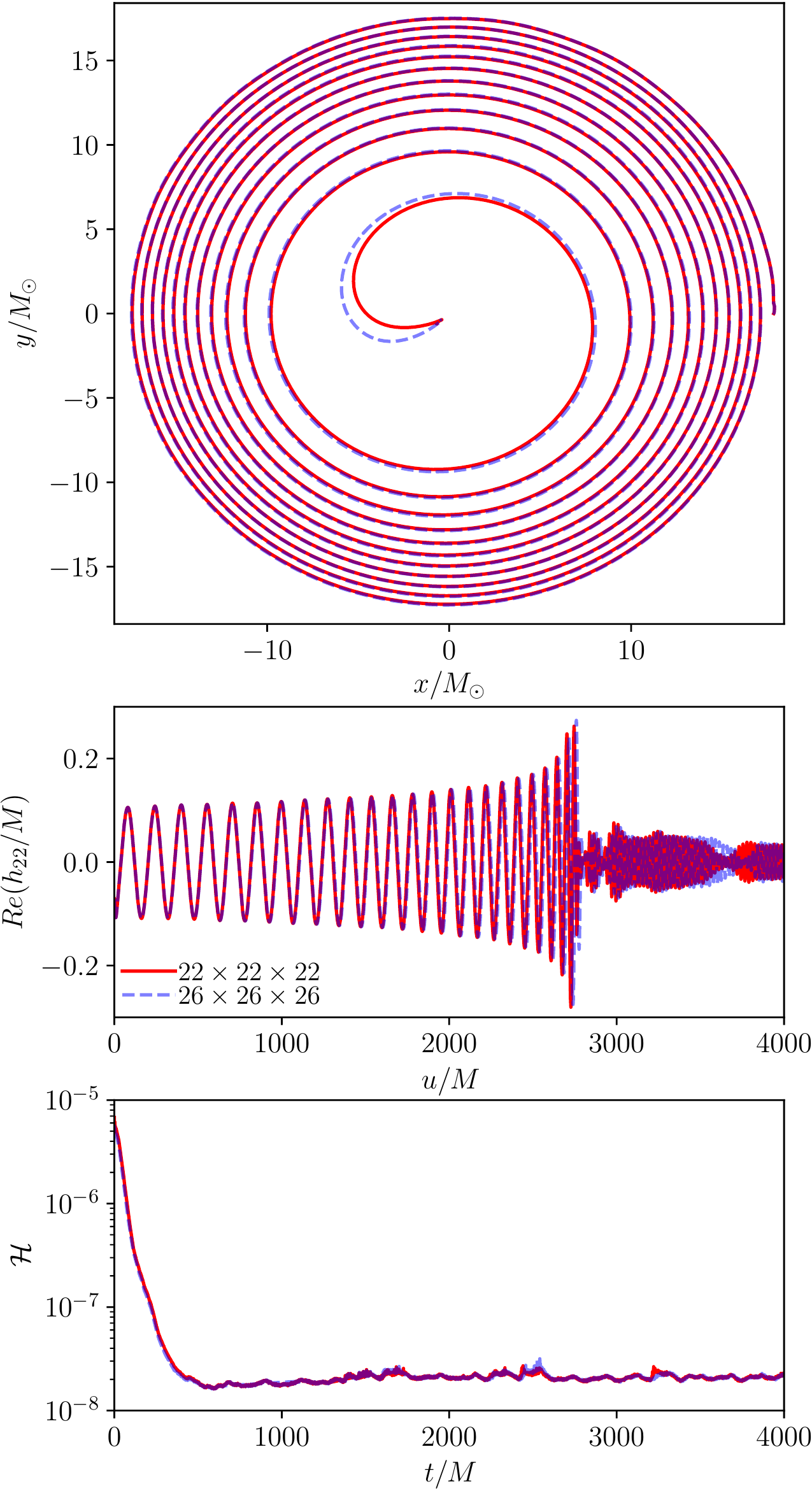}
\caption{
NS tracks of a binary pulsar system as described in Tab.~\ref{tab:millisecond}
for two different resolutions (top panel).
Real part of the dominant (2,2)-mode of the GW signal ($Re(h_{22})$)
for both resolutions (middle panel) and Hamiltonian constraint (bottom panel).
\label{fig:Binary_pulsar} }
\end{figure}

As discussed in the introduction, NSs are expected to be spinning and a
number of millisecond pulsars have been observed already (although none of
them bound in a BNS system). To proof that our upgraded SGRID version is
capable of simulating millisecond pulsars, we will present an equal mass,
aligned spin configuration in which the individual baryonic masses of the
two stars are $m_{1,2}^b=1.494607$ and the rotational velocity,
Eq.~(\ref{w-choice}), is set to $\omega_{1,2}=0.03$.

We compute initial configurations for this system with two different SGRID
resolutions, using $22\times 22 \times 22$ and $26\times 26 \times 26$
points in all domains. While the lower resolution result for this
challenging configuration can be computed in $52.4$ hours, the higher
resolution run takes about $93.2$ hours. Both initial data computations were
performed on a single Intel Xeon node with 20 cores on FAU's Koko cluster.
Due to the different resolutions, the initial configurations are slightly
different, as shown in Tab.~\ref{tab:millisecond}. We find differences
within the estimated masses of about $2\%$ and dimensionless spins of about
$3\%$ between the quasi-local mass/spin measure (Sec.~\ref{localQuantities})
and the single star properties of a NS with the same EOS, baryonic mass, and
rotational velocity. 
These differences show that the introduced quasi-local mass measure allows
only an approximate extraction of the individual masses for binary
configurations. For a high-quality analysis of high-resolution data $2\%$
differences in the individual masses, i.e., absolute differences of the
order of $\sim 10^{-3}$, are well above the acceptable uncertainty of an
analysis of the energetics of the system for which uncertainties of
$\sim 10^{-5}$ are typically required; see
e.g.~\cite{Damour:2011fu,Bernuzzi:2013rza,Dietrich:2017feu}. We thus
recommend to use the ADM mass of a single star with the same baryonic mass
and spin as the best available measure for the mass of an individual star.
However, the situation is different for the introduced quasi-local spin
measure. The fact that there is a $3\%$ difference between the quasi-local
spin of a star in a binary and the spin of a single star with the same EOS,
baryonic mass, and rotational velocity, does not mean that the quasi-local
spin measure has a $3\%$ error. Rather it is quite likely that we are simply
comparing two stars with different spins, because using the same rotational
velocity ($\omega=0.03$) does not necessarily lead to the same spin when we
compare a star in a binary and a single star.

Despite these small differences, each case describes a binary in which both
stars spin close to break-up. As far as we know, this is the highest
spinning BNS simulation that includes the merger and postmerger, which
has been performed until now. We evolve the system with the BAM code using
$96$ points within the finest refinement level. This resolution is not
sufficient for a highly accurate GW signal needed for waveform model
development, but sufficient to show that the simulation of binary
millisecond systems is feasible. The NS tracks (for one star), the emitted
GW signal, and the Hamiltonian constraint for the two resolutions are shown
in Fig.~\ref{fig:Binary_pulsar}. We find almost circular orbits
with a residual eccentricity of $\sim 10^{-3}$, due to the
employed eccentricity reduction. The difference between the phases of the GW
signals shown in the middle panel of Fig.~\ref{fig:Binary_pulsar} is about
$1$ radian at the moment of merger.
It is caused by (i) the different resolutions of
the initial data, (ii) the slightly different masses of the configurations
(cf.~Tab.~\ref{tab:millisecond}), and (iii) by the fact that
eccentricity reduction was only applied to the low SGRID resolution,
while simply using the same values for $v_r$ and $\Omega$ for the
high SGRID resolution. The bottom panel shows the Hamiltonian constraints,
where we find only minor differences between the two SGRID resolutions.

%%%%%%%%%%%%%%%%%%%%%%%%%%%%%%%%%%%%%%%%%%%%%%%%%%%%%
\subsection{Evolving highly compact stars}

In the past we had implemented the Hamiltonian constraint as
in Eq.~(\ref{ham}) and found that we were able to find a solution
only for low compactness. With the modification given by
Eq.~(\ref{hamMOD}) and described in Sec.~\ref{ConFacMod} we can now
construct much more compact stars.
As an example we have considered an equal mass binary without spin, where
each star has a baryonic mass of $m_0 = 2.4$ and obeys the SLy equation of
state. This baryonic mass corresponds to a gravitational mass of
$m= 2.0213$ and a compactness of
$\left(\frac{m}{R}\right)_{\infty}=0.284$ for each star at infinite
separation. The gravitational mass is thus very close to the maximum
$m_{max}=2.0606$ possible with the SLy equation of state. As far as we know,
it is also the most compact BNS system evolved so far.

We have evolved this binary with BAM using a piecewise-polytropic fit for
the SLy EOS~\cite{Read:2008iy,Dietrich:2015pxa} with an added thermal
contribution to the pressure following a $\Gamma$-law with $\Gamma=1.75$.
\begin{figure}[t]
\includegraphics[width=\columnwidth]
{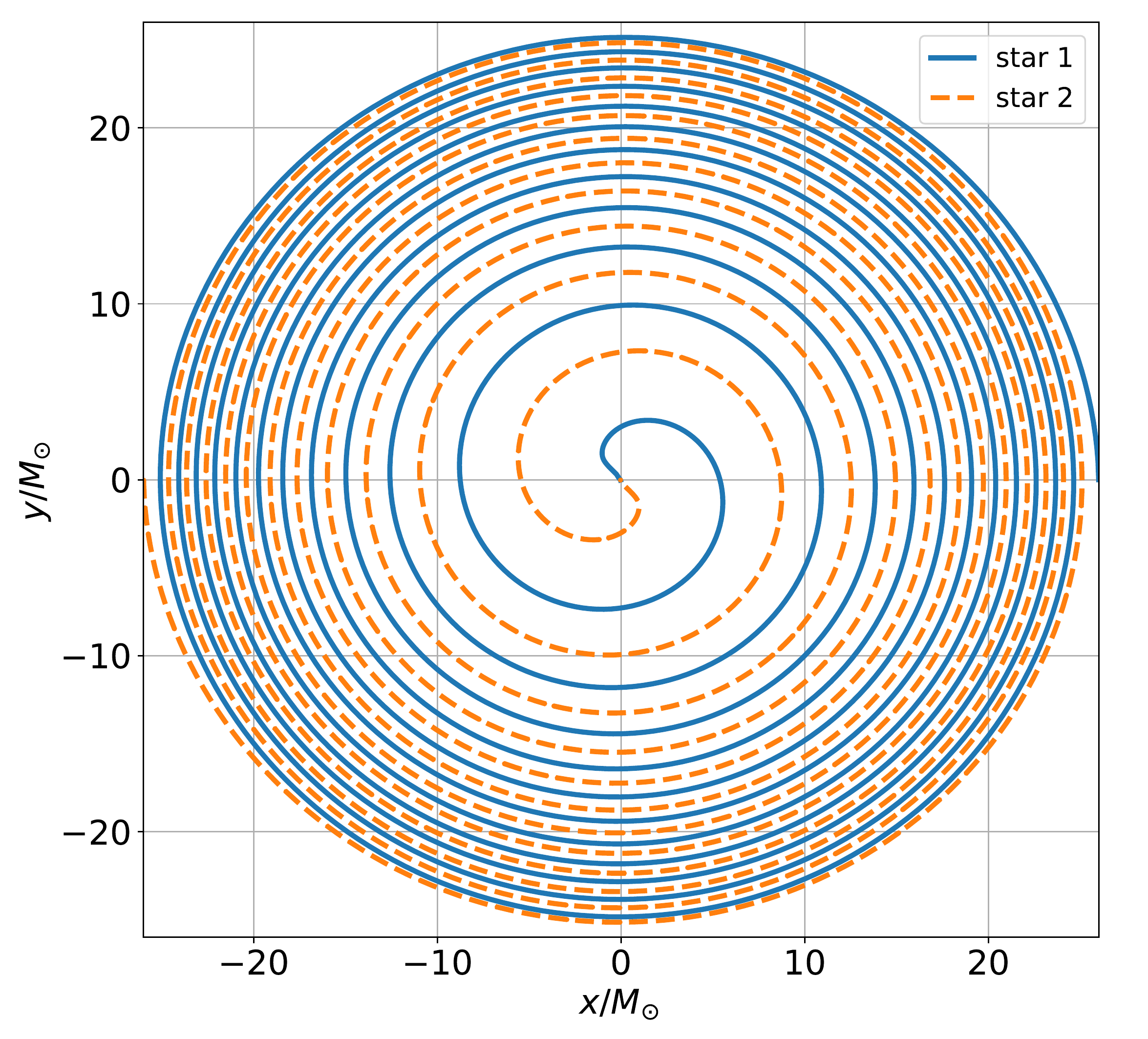}
\caption{
Tracks of the star centers for the equal mass binary with
compactness $\left(\frac{m}{R}\right)_{\infty}=0.284$.
\label{fig:High_compactness}
}
\end{figure}
In Fig.~\ref{fig:High_compactness} we show the tracks of the two star
centers starting from an inital coordinate separation of $52$ up to merger.
The initial orbital angular velocity and radial velocity are
$\Omega=0.0048738$ and $v_r=-0.00151$. The latter values have been obtained
using the eccentricity reduction procedure described in
Appendix~\ref{EccRed-Proc}.

%%%%%%%%%%%%%%%%%%%%%%%%%%%%%%%%%%%%%%%%%%%%%%%%%%%%%
\subsection{Evolving unequal mass systems}

\begin{figure}[ht]
\includegraphics[width=\columnwidth]{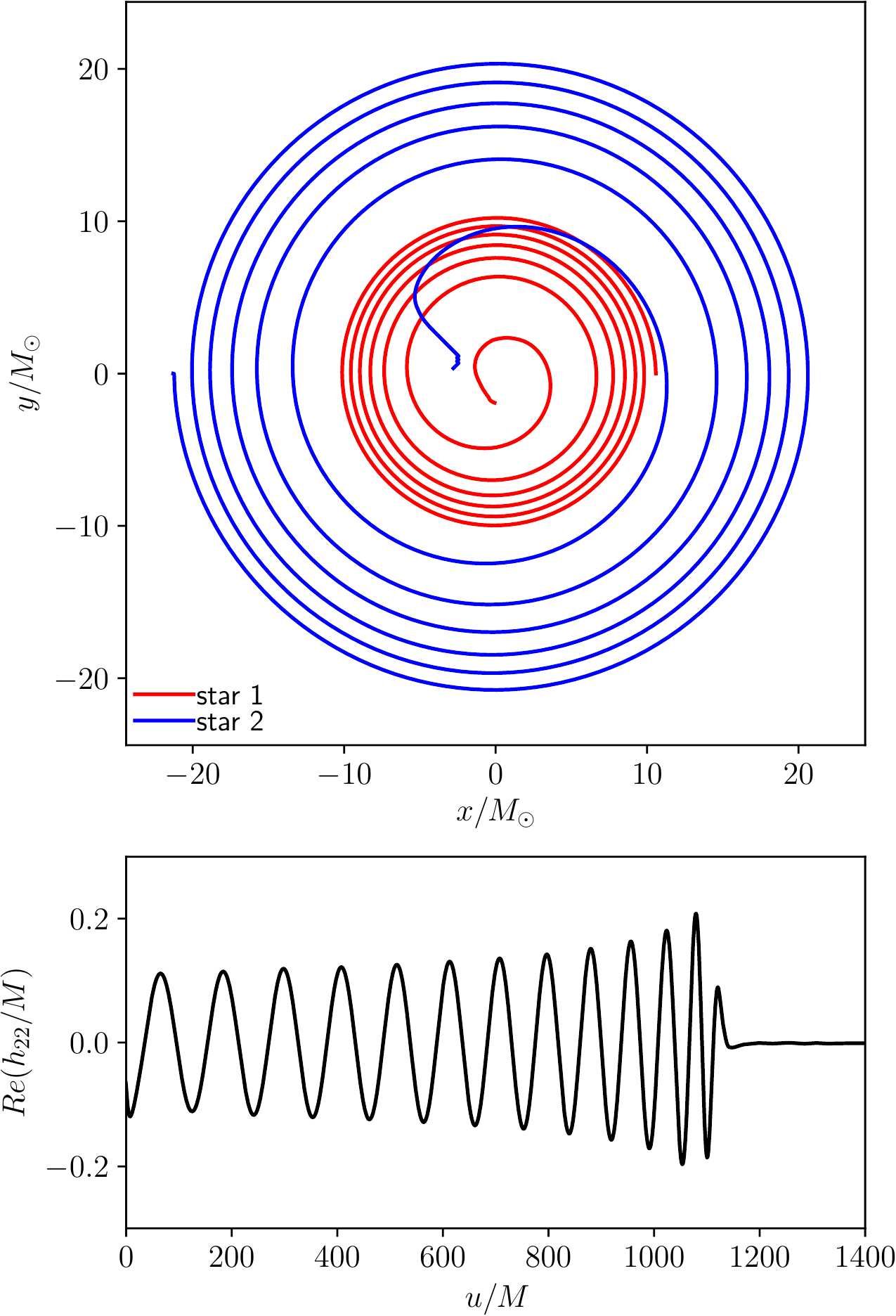}
\caption{
NS tracks of the two stars in our high-mass ratio simulation (top panel) and the
real part of the dominant (2,2)-mode of the GW signal (bottom panel).
\label{fig:High_mass_ratio} }
\end{figure}

In order to cover a larger set of configurations for binary neutron stars
and to test the capability of the new version of SGRID,
we have also constructed the initial data for a high mass ratio system.
We chose the configuration to be composed of
two non-spinning neutron stars with a piecewise-polytropic fit of the
SLy EOS~\cite{Read:2008iy,Dietrich:2015pxa} with gravitational mass of
$1.99 M_{\odot}$ and $0.98 M_{\odot}$ which results in a mass ratio of
$q = 2.03$.
This is the highest mass ratio considered for a soft equation of state in numerical
relativity for a BNS system.
While this mass ratios might even be at the edge of what is theoretically
allowed, a study of these kind of systems is essential to develop and improve
waveform models, see e.g.~\cite{Dietrich:2019kaq}.

In Fig.~\ref{fig:High_mass_ratio} we show the
tracks of each neutron star in the binary after three steps
of eccentricity reduction. These
tracks illustrate the trajectory of center of each neutron star in x-y plane.
The center of each neutron star is estimated as
the minimum of the lapse inside each star. Near merger,
the less massive star is disrupted, which causes the track of the
less massive star to end.

In Fig.~\ref{fig:High_mass_ratio} we show
the dominant (2,2)-mode of the GW ($Re(h_{22})$) versus the retarded time.
Due to the very large mass of the primary star the system undergoes
a prompt collapse to a BH after the moment of merger.
The gravitational wave signal thus settles down very quickly
after the merger.

%%%%%%%%%%%%%%%%%%%%%%%%%%%%%%%%%%%%%%%%%%
\section{Summary}
\label{summary}
%%%%%%%%%%%%%%%%%%%%%%%%%%%%%%%%%%%%%%%%%%

In this article, we have presented upgrades made to the SGRID code
to improve the capability of constructing initial data for numerical
relativity simulations.
Among other things our upgrades involve a new grid structure, the
use of different coordinates, as well as a reformulation of the
equations for the conformal factor and the velocity potential.
In order to compare with other methods or models, for example post-Newtonian
theory, one would like to know certain physical quantities such as the mass
and spin of each star. We have presented simple estimates for the initial
mass, spin, momentum, and center of mass of each individual star.

We have tested our new implementation by comparing results against the
previous SGRID version and found good agreement between initial data
sequences. We also observe lower constraint violations
(see Appendix~\ref{old_vs_new_sgrid}), and in addition
are able to construct more demanding initial data sets with high spins,
masses, and mass ratios.

To show that the new code version will be of importance within the field of
numerical relativity, we have constructed initial data for a binary
system with individual stars close to the breakup, as well as close to the
maximum mass allowed by the equation of state, and furthermore a BNS system
with a soft equation of state characterized by a high mass
ratio of $q=2.03$. All these simulations enter
previously unexplored regions of the BNS parameter space. Due to an
eccentricity reduction procedure, the presented simulations have typical
eccentricities of $\sim 10^{-3}$. This allows their usage for the calibration
and validation of gravitational waveform models.

In the future, we plan to use SGRID's new capabilities to perform new
simulations and extend the publicly available CoRe
database~\cite{Dietrich:2018phi} with high quality data, previously
not accessible within the numerical relativity community.

\begin{acknowledgments}

It is a pleasure to thank Sebastiano Bernuzzi, Erik Lundberg,
and Jason Mireles-James for helpful discussions.
This work was supported by NSF grant PHY-1707227 and DFG grant 
BR 2176/5-1.
Tim Dietrich acknowledges support by the European Union's Horizon 2020 
research and innovation program under grant agreement No 749145, BNSmergers.
We also acknowledge usage of computer time on the
HPC cluster KOKO at Florida Atlantic University, 
on the Minerva cluster at the Max Planck Institute 
for Gravitational Physics, 
on SuperMUC at the LRZ (Munich) under the project 
number pn56zo, and on the ARA cluster of 
the University of Jena.
\end{acknowledgments}

%%%%%%%%%%%%%%%%%%%%%%%%%%%%%%%%%%%%%%%%%%%%%%%%%%
\appendix
%%%%%%%%%%%%%%%%%%%%%%%%%%%%%%%%%%%%%%%%%%%%%%%%%%

\section{Empirical $\omega$ - $\chi$ relation}

\begin{figure}[ht]
  \includegraphics[scale=0.5,clip=true]{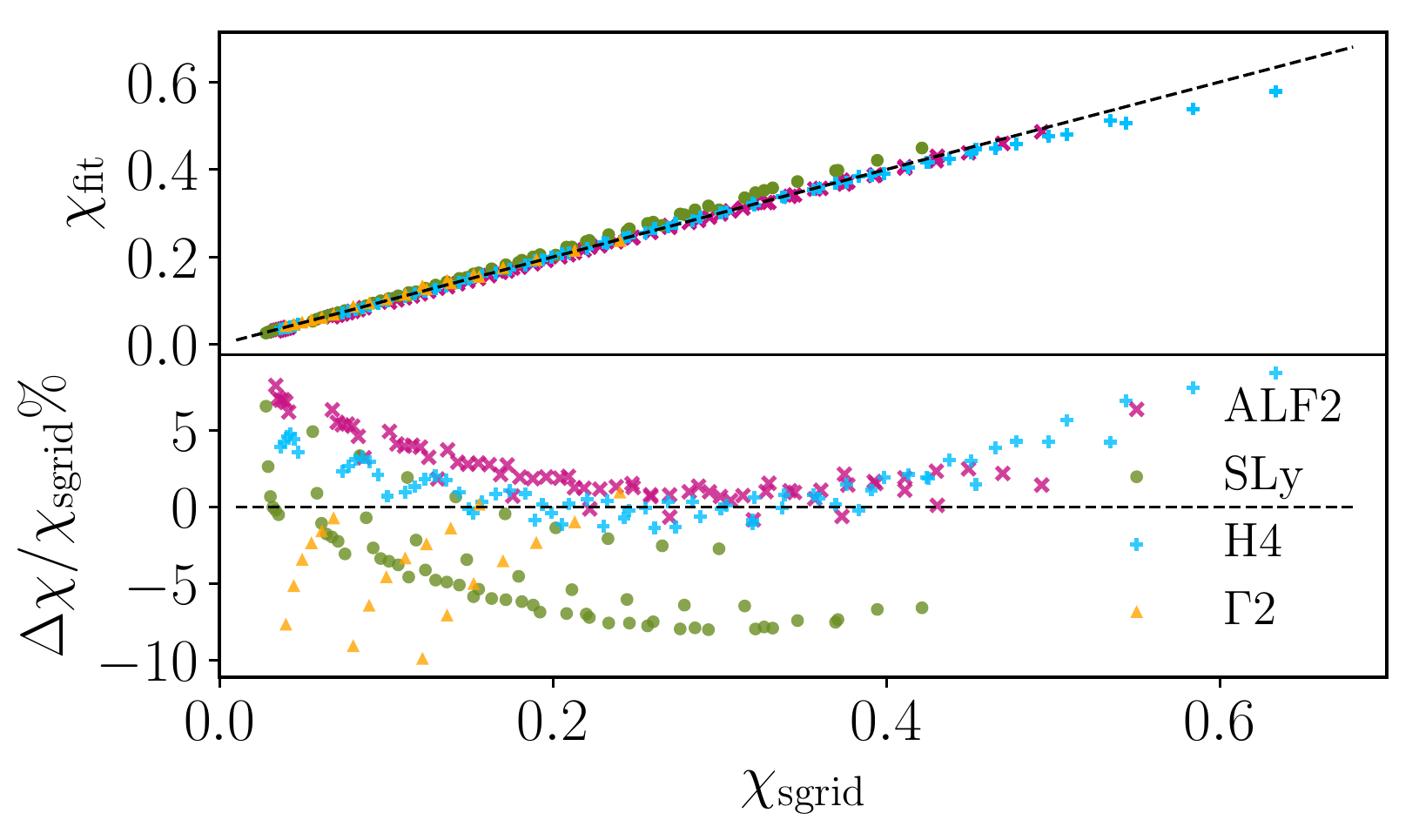}
  \caption{The dimensionless spin according to Eq.~\ref{equ:aomgfit} as a
function of spin computed in the new SGRID for different EOSs (top panel). 
The fraction residuals are shown in the bottom panel. The black dash curve 
represents $\chi_{\rm SGRID} =\chi_{\rm fit}$ scenario.
   \label{fig:a-omgfit}}
\end{figure}

As shown, SGRID can construct initial configurations in which the individual stars are arbitrarily
spinning~\cite{Dietrich:2015pxa, Tichy:2006qn,  Tichy:2009yr}. 
However, for this, one has to specify the angular velocity of the fluid 
$\omega$, the baryonic masses, and the EOS as input parameters. 
The spin itself can not be specified directly. 
Thus, to minimize computational costs and simplify the computation, 
we need to find an ansatz for the spin in terms of SGRID's input parameter. 
One such phenomenological fit has been given in Appendix C.2 of 
Ref.~\cite{Dietrich:2015pxa}. However, we found that it might give large 
errors at high spins, which are now reachable with our new SGRID implementation. 
Therefore, building upon that, we fit the following data
generated for a single star to the SGRID output for $\chi_{\rm SGRID}$. 
We use 4 EOSs, (SLy, ALF2, H4, and a $\Gamma=2$ polytrope) with baryonic masses 
$M_b/M_\odot \in [1.1, 1.7]$ in steps of $0.1$ and compactnesses in 
the range of $\mathcal{C} \in [0.09, 0.20]$. 
We find the following phenomenological fit for the dimensionless 
spin magnitude $\chi$ of a single NS: 
\begin{equation}
 \chi_{\rm fit} = a_1( 1+ m_1M_b)(1  +c_1\mathcal{C} + c_2\mathcal{C}^2 +c_3\mathcal{C}^3 +  c_4\mathcal{C}^4)(1 + d_1 \omega)\omega,
 \label{equ:aomgfit}
\end{equation}
where the coefficients $a_1 = 59.329$, $m_1 = 1.9267$, $c_1 = -17.1537$, 
$c_2 = 122.8986$, $c_3 =-401.3542$, $c_4 = 483.0869$, and $d_1 = 10.2497$ 
are computed by fitting the data, cf.\ Fig.~\ref{fig:a-omgfit}. 
Specifically, we employ for all combinations of the NS mass and EOS, ten different values 
of $\omega \in [0.000, 0.02]$ in steps of $\Delta \omega = 0.002$. 
The fractional residuals for each configuration is shown in the bottom panel 
Fig.~\ref{fig:a-omgfit}. The new fit gives maximum 10\% error 
for some extreme cases otherwise the error is below 5\%.

%%%%%%%%%%%%%%%%%%%%%%%%%%%%%%%%%%%%%%%%%%%%%%%%%%%%%
\section{$\Omega$ based eccentricity reduction procedure}
\label{EccRed-Proc}

In most cases we have used an eccentricity reduction procedure very similar
to the one in~\cite{Tacik:2015tja}, instead of the one described
in~\cite{Dietrich:2015pxa}, because in many cases it is advantageous to
avoid using the ``force balance'' relation mentioned in point~\ref{it-Omega}
of Sec.~(\ref{it-scheme}).

We start with a post-Newtonian estimate for $\Omega$ as well as $v_r=0$. We
then evolve for about three orbits and fit the observed distance $d(t)$
between the star centers to
\be
S(t) = S_0 + A_0 t + \frac{A_1}{2} t^2
       - \frac{B}{\omega_f}\cos(\omega_f t + \phi) ,
\ee
where $S_0$, $A_0$, $A_1$, $B$, $\omega_f$, and $\phi$ are fit parameters.
From the fit parameters we compute the measured eccentricity
\be
e = -\frac{B}{\omega_f d_0}
\ee
as well as the changes
\be
\delta v_r = -B \sin\phi, \ \ \
\delta\Omega = -\frac{B\omega_f\cos\phi}{2\Omega d_0}
\ee
in $v_r$ and $\Omega$ needed to lower the eccentricity. We then recompute
initial data with the thus changed values for $v_r$ and $\Omega$,
and evolve and fit again to obtain the next set of changes to $v_r$ and
$\Omega$. We usually perform 3 or 4 such reduction steps. Notice that we
typically use the proper distance as the distance measure $d(t)$ that we
fit, and that we set $d_0$ equal to the initial coordinate distance.
The latter has given slightly better estimates for $\delta v_r$ and
$\delta\Omega$ than simply setting $d_0=S_0$.

%%%%%%%%%%%%%%%%%%%%%%%%%%%%%%%%%%%%%%%%%%%%%%%%%%%%%
\section{Comparison with the old version of SGRID}
\label{old_vs_new_sgrid}

\begin{figure}[t]
\includegraphics[scale=0.4,clip=true]
{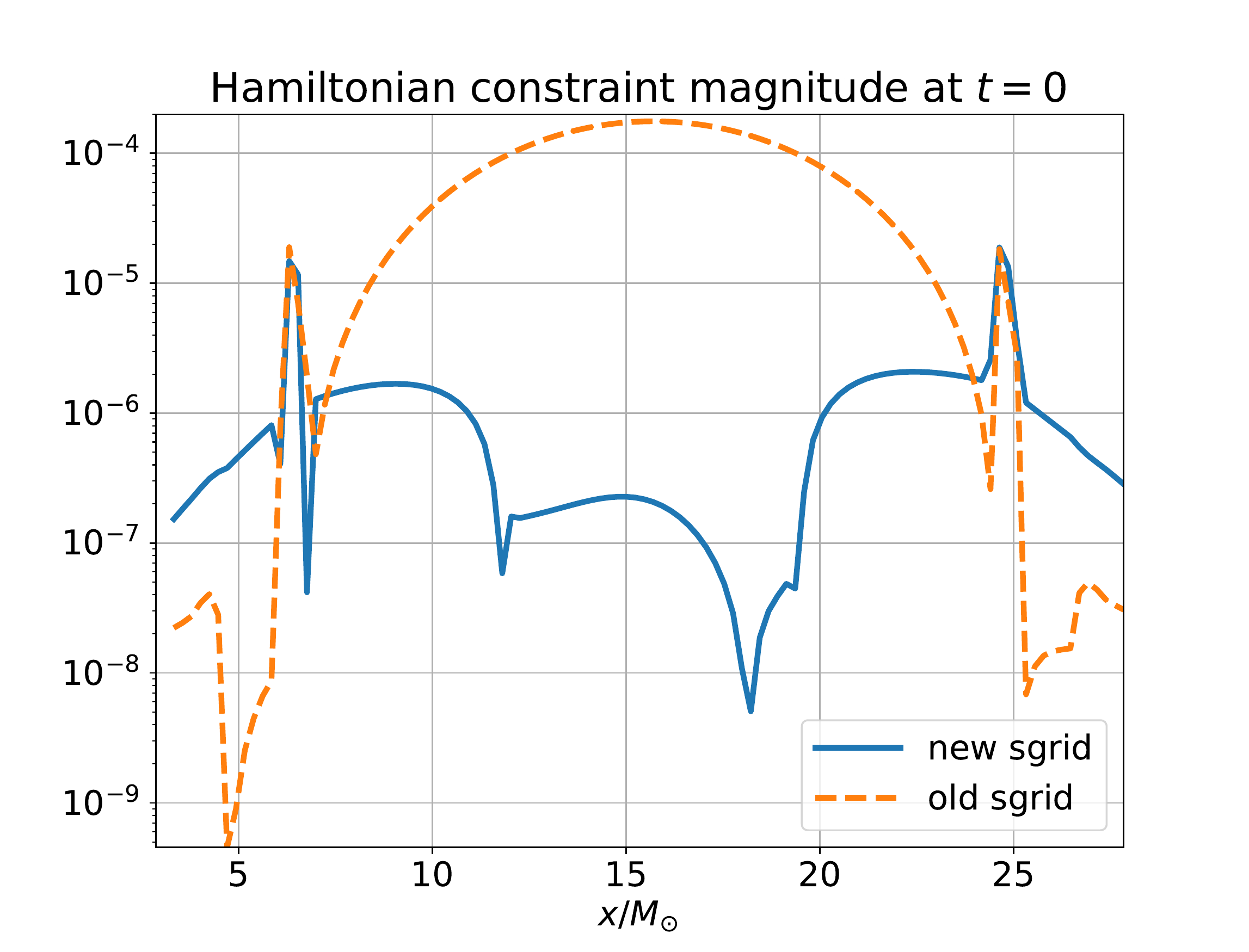}
\caption{
\label{fig:ham_vs_x}
The Hamiltonian constraint across one of the stars at the initial time
for both the old (broken line) and new (solid line) version of SGRID.
Similar violations occur in both approaches near the star surfaces,
but inside the stars the new version shows less violations.
}
\end{figure}

\begin{figure}[t]
\includegraphics[scale=0.49,clip=true]{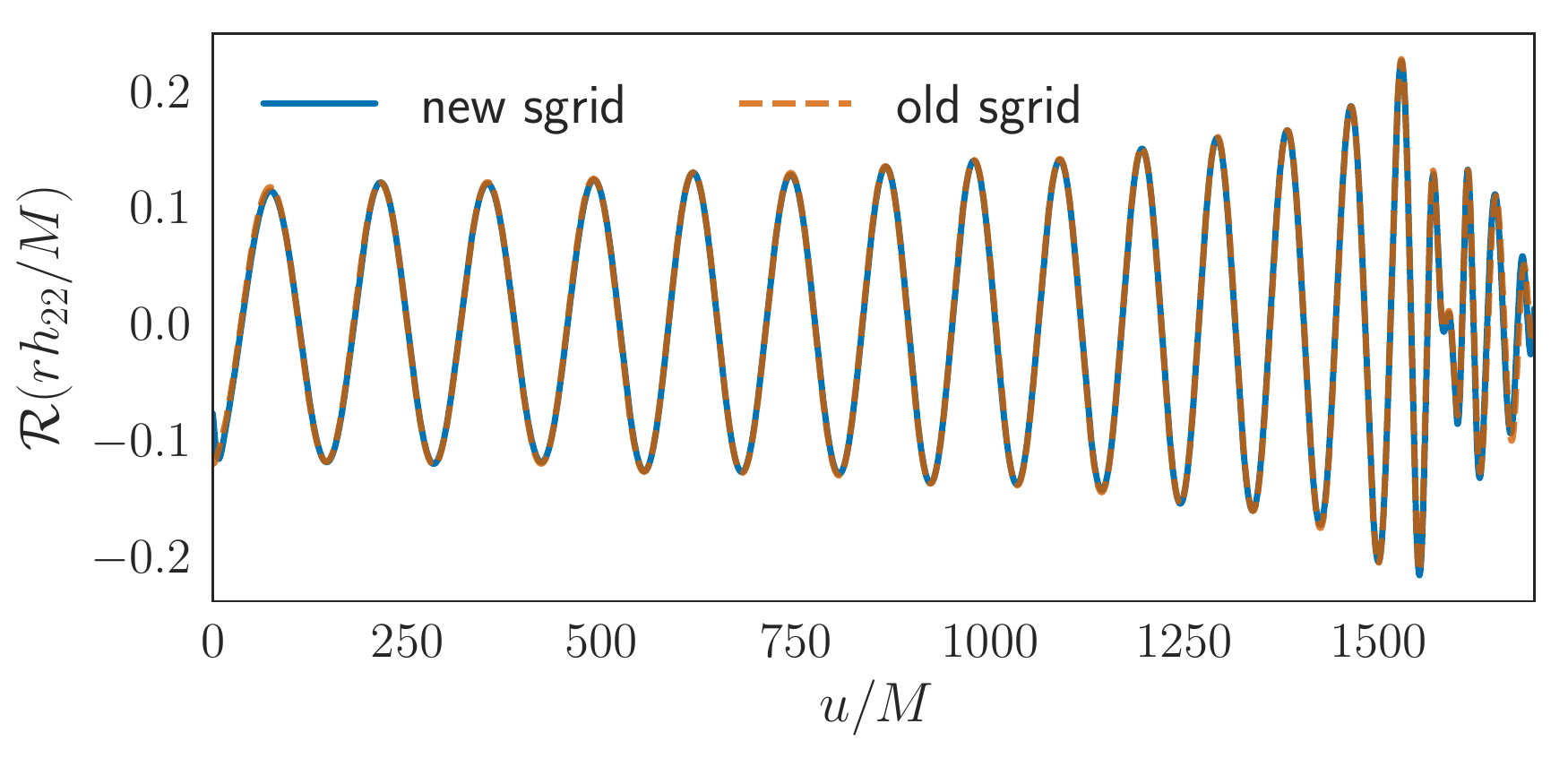}
\caption{\label{fig:SGRID_old-new_waves}
Comparison between the GW signals computed for the same physical configurations 
computed with the old and new SGRID code. 
Old SGRID results are shown with an orange dashed line, 
new SGRID results with a blue, solid curve. }
\end{figure}

In order to test the new implementation, we have constructed and evolved
initial data with the same physical parameters using the
two different SGRID versions.
We use the same configuration for both initial data, namely a $\Gamma=2$,
$\kappa= 123.6489$ EOS. The system is an equal mass binary in which the
individual stars have a baryonic mass of 1.625 $M_\odot$. The initial
separation between the stars is $68.8\rm km$.

Figure~\ref{fig:ham_vs_x} shows the Hamiltonian constraint across one of the
stars at the initial time after interpolating the SGRID data onto BAM's
grid. As we can see the new SGRID version (solid
line) produces smaller constraint violations than the old version (broken
line), inside the star, while at the star surfaces both lead to
approximately the same violations. 
Outside the stars, the old SGRID version seems slightly superior. 

In Fig.~\ref{fig:SGRID_old-new_waves} we show the dominant (2,2) mode of GW.  
We evolve both initial data sets with BAM using exactly the same setup for both evolution, namely 
6 refinement and 96 points to cover the star. 
The GWs are extracted at a distance of 900 $M_\odot$. 
Waveforms are aligned for the two cases at early times, 
i.e., before $u \leq 600 M$. 
We find that both waves agree very well throughout the merger 
and in the early post-merger part; see Fig.~\ref{fig:SGRID_old-new_waves}.

%%%%%%%%%%%%%%%%%%%%%%%%%%%%%%%%%%%%%%%%%%%%%%%%%%

%%%%%%%%%%%%%%%%%%%%%%%%%%%%%%%%%%%%%%%%%%%%%%%

%%%%%%%%%%%%%%%%%%%%%%%%%%%%%%%%%%%%%%%%%%%%%%%
% REFERENCES
%%%%%%%%%%%%%%%%%%%%%%%%%%%%%%%%%%%%%%%%%%%%%%%
\bibliography{references}

\begin{thebibliography}{76}
\expandafter\ifx\csname natexlab\endcsname\relax\def\natexlab#1{#1}\fi
\expandafter\ifx\csname bibnamefont\endcsname\relax
  \def\bibnamefont#1{#1}\fi
\expandafter\ifx\csname bibfnamefont\endcsname\relax
  \def\bibfnamefont#1{#1}\fi
\expandafter\ifx\csname citenamefont\endcsname\relax
  \def\citenamefont#1{#1}\fi
\expandafter\ifx\csname url\endcsname\relax
  \def\url#1{\texttt{#1}}\fi
\expandafter\ifx\csname urlprefix\endcsname\relax\def\urlprefix{URL }\fi
\providecommand{\bibinfo}[2]{#2}
\providecommand{\eprint}[2][]{\url{#2}}

\bibitem[{\citenamefont{Abbott
  et~al.}(2017{\natexlab{a}})}]{TheLIGOScientific:2017qsa}
\bibinfo{author}{\bibfnamefont{B.~P.} \bibnamefont{Abbott}}
  \bibnamefont{et~al.} (\bibinfo{collaboration}{LIGO Scientific, Virgo}),
  \bibinfo{journal}{Phys. Rev. Lett.} \textbf{\bibinfo{volume}{119}},
  \bibinfo{pages}{161101} (\bibinfo{year}{2017}{\natexlab{a}}),
  \eprint{1710.05832}.

\bibitem[{\citenamefont{Abbott et~al.}(2017{\natexlab{b}})}]{GBM:2017lvd}
\bibinfo{author}{\bibfnamefont{B.~P.} \bibnamefont{Abbott}}
  \bibnamefont{et~al.} (\bibinfo{collaboration}{LIGO Scientific, Virgo, Fermi
  GBM, INTEGRAL, IceCube, AstroSat Cadmium Zinc Telluride Imager Team, IPN,
  Insight-Hxmt, ANTARES, Swift, AGILE Team, 1M2H Team, Dark Energy Camera
  GW-EM, DES, DLT40, GRAWITA, Fermi-LAT, ATCA, ASKAP, Las Cumbres Observatory
  Group, OzGrav, DWF (Deeper Wider Faster Program), AST3, CAASTRO, VINROUGE,
  MASTER, J-GEM, GROWTH, JAGWAR, CaltechNRAO, TTU-NRAO, NuSTAR, Pan-STARRS,
  MAXI Team, TZAC Consortium, KU, Nordic Optical Telescope, ePESSTO, GROND,
  Texas Tech University, SALT Group, TOROS, BOOTES, MWA, CALET, IKI-GW
  Follow-up, H.E.S.S., LOFAR, LWA, HAWC, Pierre Auger, ALMA, Euro VLBI Team, Pi
  of Sky, Chandra Team at McGill University, DFN, ATLAS Telescopes, High Time
  Resolution Universe Survey, RIMAS, RATIR, SKA South Africa/MeerKAT}),
  \bibinfo{journal}{Astrophys. J.} \textbf{\bibinfo{volume}{848}},
  \bibinfo{pages}{L12} (\bibinfo{year}{2017}{\natexlab{b}}),
  \eprint{1710.05833}.

\bibitem[{\citenamefont{Cook}(2000)}]{Cook00a}
\bibinfo{author}{\bibfnamefont{G.~B.} \bibnamefont{Cook}},
  \bibinfo{journal}{Living Rev. Rel.} \textbf{\bibinfo{volume}{3}},
  \bibinfo{pages}{5} (\bibinfo{year}{2000}), \eprint{gr-qc/0007085}.

\bibitem[{\citenamefont{Tichy}(2017)}]{Tichy:2016vmv}
\bibinfo{author}{\bibfnamefont{W.}~\bibnamefont{Tichy}},
  \bibinfo{journal}{Rept. Prog. Phys.} \textbf{\bibinfo{volume}{80}},
  \bibinfo{pages}{026901} (\bibinfo{year}{2017}), \eprint{1610.03805}.

\bibitem[{\citenamefont{Haas et~al.}(2016)}]{Haas:2016cop}
\bibinfo{author}{\bibfnamefont{R.}~\bibnamefont{Haas}} \bibnamefont{et~al.},
  \bibinfo{journal}{Phys. Rev.} \textbf{\bibinfo{volume}{D93}},
  \bibinfo{pages}{124062} (\bibinfo{year}{2016}), \eprint{1604.00782}.

\bibitem[{\citenamefont{Favata}(2014)}]{Favata:2013rwa}
\bibinfo{author}{\bibfnamefont{M.}~\bibnamefont{Favata}},
  \bibinfo{journal}{Phys. Rev. Lett.} \textbf{\bibinfo{volume}{112}},
  \bibinfo{pages}{101101} (\bibinfo{year}{2014}), \eprint{1310.8288}.

\bibitem[{\citenamefont{Agathos et~al.}(2015)\citenamefont{Agathos, Meidam,
  Del~Pozzo, Li, Tompitak, Veitch, Vitale, and Broeck}}]{Agathos:2015uaa}
\bibinfo{author}{\bibfnamefont{M.}~\bibnamefont{Agathos}},
  \bibinfo{author}{\bibfnamefont{J.}~\bibnamefont{Meidam}},
  \bibinfo{author}{\bibfnamefont{W.}~\bibnamefont{Del~Pozzo}},
  \bibinfo{author}{\bibfnamefont{T.~G.~F.} \bibnamefont{Li}},
  \bibinfo{author}{\bibfnamefont{M.}~\bibnamefont{Tompitak}},
  \bibinfo{author}{\bibfnamefont{J.}~\bibnamefont{Veitch}},
  \bibinfo{author}{\bibfnamefont{S.}~\bibnamefont{Vitale}}, \bibnamefont{and}
  \bibinfo{author}{\bibfnamefont{C.~V.~D.} \bibnamefont{Broeck}},
  \bibinfo{journal}{Phys. Rev.} \textbf{\bibinfo{volume}{D92}},
  \bibinfo{pages}{023012} (\bibinfo{year}{2015}), \eprint{1503.05405}.

\bibitem[{\citenamefont{Samajdar and Dietrich}(2019)}]{Samajdar:2019ulq}
\bibinfo{author}{\bibfnamefont{A.}~\bibnamefont{Samajdar}} \bibnamefont{and}
  \bibinfo{author}{\bibfnamefont{T.}~\bibnamefont{Dietrich}}
  (\bibinfo{year}{2019}), \eprint{1905.03118}.

\bibitem[{\citenamefont{Hessels et~al.}(2006)\citenamefont{Hessels, Ransom,
  Stairs, Freire, Kaspi, and Camilo}}]{Hessels:2006ze}
\bibinfo{author}{\bibfnamefont{J.~W.~T.} \bibnamefont{Hessels}},
  \bibinfo{author}{\bibfnamefont{S.~M.} \bibnamefont{Ransom}},
  \bibinfo{author}{\bibfnamefont{I.~H.} \bibnamefont{Stairs}},
  \bibinfo{author}{\bibfnamefont{P.~C.~C.} \bibnamefont{Freire}},
  \bibinfo{author}{\bibfnamefont{V.~M.} \bibnamefont{Kaspi}}, \bibnamefont{and}
  \bibinfo{author}{\bibfnamefont{F.}~\bibnamefont{Camilo}},
  \bibinfo{journal}{Science} \textbf{\bibinfo{volume}{311}},
  \bibinfo{pages}{1901} (\bibinfo{year}{2006}), \eprint{astro-ph/0601337}.

\bibitem[{\citenamefont{Lynch et~al.}(2012)\citenamefont{Lynch, Freire, Ransom,
  and Jacoby}}]{Lynch:2011aa}
\bibinfo{author}{\bibfnamefont{R.~S.} \bibnamefont{Lynch}},
  \bibinfo{author}{\bibfnamefont{P.~C.~C.} \bibnamefont{Freire}},
  \bibinfo{author}{\bibfnamefont{S.~M.} \bibnamefont{Ransom}},
  \bibnamefont{and} \bibinfo{author}{\bibfnamefont{B.~A.}
  \bibnamefont{Jacoby}}, \bibinfo{journal}{Astrophys. J.}
  \textbf{\bibinfo{volume}{745}}, \bibinfo{pages}{109} (\bibinfo{year}{2012}),
  \eprint{1112.2612}.

\bibitem[{\citenamefont{Stovall et~al.}(2018)}]{Stovall:2018ouw}
\bibinfo{author}{\bibfnamefont{K.}~\bibnamefont{Stovall}} \bibnamefont{et~al.},
  \bibinfo{journal}{Astrophys. J.} \textbf{\bibinfo{volume}{854}},
  \bibinfo{pages}{L22} (\bibinfo{year}{2018}), \eprint{1802.01707}.

\bibitem[{\citenamefont{Cromartie et~al.}(2019)}]{Cromartie:2019kug}
\bibinfo{author}{\bibfnamefont{H.~T.} \bibnamefont{Cromartie}}
  \bibnamefont{et~al.} (\bibinfo{year}{2019}), \eprint{1904.06759}.

\bibitem[{\citenamefont{Dietrich and Ujevic}(2017)}]{Dietrich:2016fpt}
\bibinfo{author}{\bibfnamefont{T.}~\bibnamefont{Dietrich}} \bibnamefont{and}
  \bibinfo{author}{\bibfnamefont{M.}~\bibnamefont{Ujevic}},
  \bibinfo{journal}{Class. Quant. Grav.} \textbf{\bibinfo{volume}{34}},
  \bibinfo{pages}{105014} (\bibinfo{year}{2017}), \eprint{1612.03665}.

\bibitem[{\citenamefont{Bauswein et~al.}(2013)\citenamefont{Bauswein,
  Baumgarte, and Janka}}]{Bauswein:2013jpa}
\bibinfo{author}{\bibfnamefont{A.}~\bibnamefont{Bauswein}},
  \bibinfo{author}{\bibfnamefont{T.~W.} \bibnamefont{Baumgarte}},
  \bibnamefont{and} \bibinfo{author}{\bibfnamefont{H.~T.} \bibnamefont{Janka}},
  \bibinfo{journal}{Phys. Rev. Lett.} \textbf{\bibinfo{volume}{111}},
  \bibinfo{pages}{131101} (\bibinfo{year}{2013}), \eprint{1307.5191}.

\bibitem[{\citenamefont{Köppel et~al.}(2019)\citenamefont{Köppel, Bovard, and
  Rezzolla}}]{Koppel:2019pys}
\bibinfo{author}{\bibfnamefont{S.}~\bibnamefont{Köppel}},
  \bibinfo{author}{\bibfnamefont{L.}~\bibnamefont{Bovard}}, \bibnamefont{and}
  \bibinfo{author}{\bibfnamefont{L.}~\bibnamefont{Rezzolla}},
  \bibinfo{journal}{Astrophys. J.} \textbf{\bibinfo{volume}{872}},
  \bibinfo{pages}{L16} (\bibinfo{year}{2019}), \eprint{1901.09977}.

\bibitem[{\citenamefont{Hinderer et~al.}(2010)\citenamefont{Hinderer, Lackey,
  Lang, and Read}}]{Hinderer:2009ca}
\bibinfo{author}{\bibfnamefont{T.}~\bibnamefont{Hinderer}},
  \bibinfo{author}{\bibfnamefont{B.~D.} \bibnamefont{Lackey}},
  \bibinfo{author}{\bibfnamefont{R.~N.} \bibnamefont{Lang}}, \bibnamefont{and}
  \bibinfo{author}{\bibfnamefont{J.~S.} \bibnamefont{Read}},
  \bibinfo{journal}{Phys. Rev.} \textbf{\bibinfo{volume}{D81}},
  \bibinfo{pages}{123016} (\bibinfo{year}{2010}), \eprint{0911.3535}.

\bibitem[{\citenamefont{Dietrich
  et~al.}(2017{\natexlab{a}})\citenamefont{Dietrich, Bernuzzi, Ujevic, and
  Tichy}}]{Dietrich:2016lyp}
\bibinfo{author}{\bibfnamefont{T.}~\bibnamefont{Dietrich}},
  \bibinfo{author}{\bibfnamefont{S.}~\bibnamefont{Bernuzzi}},
  \bibinfo{author}{\bibfnamefont{M.}~\bibnamefont{Ujevic}}, \bibnamefont{and}
  \bibinfo{author}{\bibfnamefont{W.}~\bibnamefont{Tichy}},
  \bibinfo{journal}{Phys. Rev.} \textbf{\bibinfo{volume}{D95}},
  \bibinfo{pages}{044045} (\bibinfo{year}{2017}{\natexlab{a}}),
  \eprint{1611.07367}.

\bibitem[{\citenamefont{Lehner et~al.}(2016)\citenamefont{Lehner, Liebling,
  Palenzuela, Caballero, O'Connor, Anderson, and Neilsen}}]{Lehner:2016lxy}
\bibinfo{author}{\bibfnamefont{L.}~\bibnamefont{Lehner}},
  \bibinfo{author}{\bibfnamefont{S.~L.} \bibnamefont{Liebling}},
  \bibinfo{author}{\bibfnamefont{C.}~\bibnamefont{Palenzuela}},
  \bibinfo{author}{\bibfnamefont{O.~L.} \bibnamefont{Caballero}},
  \bibinfo{author}{\bibfnamefont{E.}~\bibnamefont{O'Connor}},
  \bibinfo{author}{\bibfnamefont{M.}~\bibnamefont{Anderson}}, \bibnamefont{and}
  \bibinfo{author}{\bibfnamefont{D.}~\bibnamefont{Neilsen}},
  \bibinfo{journal}{Class. Quant. Grav.} \textbf{\bibinfo{volume}{33}},
  \bibinfo{pages}{184002} (\bibinfo{year}{2016}), \eprint{1603.00501}.

\bibitem[{\citenamefont{Zappa et~al.}(2018)\citenamefont{Zappa, Bernuzzi,
  Radice, Perego, and Dietrich}}]{Zappa:2017xba}
\bibinfo{author}{\bibfnamefont{F.}~\bibnamefont{Zappa}},
  \bibinfo{author}{\bibfnamefont{S.}~\bibnamefont{Bernuzzi}},
  \bibinfo{author}{\bibfnamefont{D.}~\bibnamefont{Radice}},
  \bibinfo{author}{\bibfnamefont{A.}~\bibnamefont{Perego}}, \bibnamefont{and}
  \bibinfo{author}{\bibfnamefont{T.}~\bibnamefont{Dietrich}},
  \bibinfo{journal}{Phys. Rev. Lett.} \textbf{\bibinfo{volume}{120}},
  \bibinfo{pages}{111101} (\bibinfo{year}{2018}), \eprint{1712.04267}.

\bibitem[{\citenamefont{Kiuchi et~al.}(2019{\natexlab{a}})\citenamefont{Kiuchi,
  Kyutoku, Shibata, and Taniguchi}}]{Kiuchi:2019lls}
\bibinfo{author}{\bibfnamefont{K.}~\bibnamefont{Kiuchi}},
  \bibinfo{author}{\bibfnamefont{K.}~\bibnamefont{Kyutoku}},
  \bibinfo{author}{\bibfnamefont{M.}~\bibnamefont{Shibata}}, \bibnamefont{and}
  \bibinfo{author}{\bibfnamefont{K.}~\bibnamefont{Taniguchi}},
  \bibinfo{journal}{Astrophys. J.} \textbf{\bibinfo{volume}{876}},
  \bibinfo{pages}{L31} (\bibinfo{year}{2019}{\natexlab{a}}),
  \eprint{1903.01466}.

\bibitem[{\citenamefont{Dominik et~al.}(2012)\citenamefont{Dominik, Belczynski,
  Fryer, Holz, Berti, Bulik, Mandel, and O'Shaughnessy}}]{Dominik:2012kk}
\bibinfo{author}{\bibfnamefont{M.}~\bibnamefont{Dominik}},
  \bibinfo{author}{\bibfnamefont{K.}~\bibnamefont{Belczynski}},
  \bibinfo{author}{\bibfnamefont{C.}~\bibnamefont{Fryer}},
  \bibinfo{author}{\bibfnamefont{D.~E.} \bibnamefont{Holz}},
  \bibinfo{author}{\bibfnamefont{E.}~\bibnamefont{Berti}},
  \bibinfo{author}{\bibfnamefont{T.}~\bibnamefont{Bulik}},
  \bibinfo{author}{\bibfnamefont{I.}~\bibnamefont{Mandel}}, \bibnamefont{and}
  \bibinfo{author}{\bibfnamefont{R.}~\bibnamefont{O'Shaughnessy}},
  \bibinfo{journal}{Astrophys. J.} \textbf{\bibinfo{volume}{759}},
  \bibinfo{pages}{52} (\bibinfo{year}{2012}), \eprint{1202.4901}.

\bibitem[{\citenamefont{Dietrich et~al.}(2015)\citenamefont{Dietrich,
  Moldenhauer, Johnson-McDaniel, Bernuzzi, Markakis, Brügmann, and
  Tichy}}]{Dietrich:2015pxa}
\bibinfo{author}{\bibfnamefont{T.}~\bibnamefont{Dietrich}},
  \bibinfo{author}{\bibfnamefont{N.}~\bibnamefont{Moldenhauer}},
  \bibinfo{author}{\bibfnamefont{N.~K.} \bibnamefont{Johnson-McDaniel}},
  \bibinfo{author}{\bibfnamefont{S.}~\bibnamefont{Bernuzzi}},
  \bibinfo{author}{\bibfnamefont{C.~M.} \bibnamefont{Markakis}},
  \bibinfo{author}{\bibfnamefont{B.}~\bibnamefont{Brügmann}},
  \bibnamefont{and} \bibinfo{author}{\bibfnamefont{W.}~\bibnamefont{Tichy}},
  \bibinfo{journal}{Phys. Rev.} \textbf{\bibinfo{volume}{D92}},
  \bibinfo{pages}{124007} (\bibinfo{year}{2015}), \eprint{1507.07100}.

\bibitem[{\citenamefont{Martinez et~al.}(2015)\citenamefont{Martinez, Stovall,
  Freire, Deneva, Jenet, McLaughlin, Bagchi, Bates, and
  Ridolfi}}]{Martinez:2015mya}
\bibinfo{author}{\bibfnamefont{J.~G.} \bibnamefont{Martinez}},
  \bibinfo{author}{\bibfnamefont{K.}~\bibnamefont{Stovall}},
  \bibinfo{author}{\bibfnamefont{P.~C.~C.} \bibnamefont{Freire}},
  \bibinfo{author}{\bibfnamefont{J.~S.} \bibnamefont{Deneva}},
  \bibinfo{author}{\bibfnamefont{F.~A.} \bibnamefont{Jenet}},
  \bibinfo{author}{\bibfnamefont{M.~A.} \bibnamefont{McLaughlin}},
  \bibinfo{author}{\bibfnamefont{M.}~\bibnamefont{Bagchi}},
  \bibinfo{author}{\bibfnamefont{S.~D.} \bibnamefont{Bates}}, \bibnamefont{and}
  \bibinfo{author}{\bibfnamefont{A.}~\bibnamefont{Ridolfi}},
  \bibinfo{journal}{Astrophys. J.} \textbf{\bibinfo{volume}{812}},
  \bibinfo{pages}{143} (\bibinfo{year}{2015}), \eprint{1509.08805}.

\bibitem[{\citenamefont{Lazarus et~al.}(2016)}]{Lazarus:2016hfu}
\bibinfo{author}{\bibfnamefont{P.}~\bibnamefont{Lazarus}} \bibnamefont{et~al.},
  \bibinfo{journal}{Astrophys. J.} \textbf{\bibinfo{volume}{831}},
  \bibinfo{pages}{150} (\bibinfo{year}{2016}), \eprint{1608.08211}.

\bibitem[{lor()}]{lorene_web}
\bibinfo{note}{{LORENE: Langage Objet pour la RElativit\'e Num\'eriquE}, {\tt
  http://www.lorene.obspm.fr}}.

\bibitem[{\citenamefont{Kyutoku et~al.}(2014)\citenamefont{Kyutoku, Shibata,
  and Taniguchi}}]{Kyutoku:2014yba}
\bibinfo{author}{\bibfnamefont{K.}~\bibnamefont{Kyutoku}},
  \bibinfo{author}{\bibfnamefont{M.}~\bibnamefont{Shibata}}, \bibnamefont{and}
  \bibinfo{author}{\bibfnamefont{K.}~\bibnamefont{Taniguchi}},
  \bibinfo{journal}{Phys. Rev.} \textbf{\bibinfo{volume}{D90}},
  \bibinfo{pages}{064006} (\bibinfo{year}{2014}), \eprint{1405.6207}.

\bibitem[{\citenamefont{East et~al.}(2012)\citenamefont{East, Ramazanoglu, and
  Pretorius}}]{East:2012zn}
\bibinfo{author}{\bibfnamefont{W.~E.} \bibnamefont{East}},
  \bibinfo{author}{\bibfnamefont{F.~M.} \bibnamefont{Ramazanoglu}},
  \bibnamefont{and}
  \bibinfo{author}{\bibfnamefont{F.}~\bibnamefont{Pretorius}},
  \bibinfo{journal}{Phys. Rev.} \textbf{\bibinfo{volume}{D86}},
  \bibinfo{pages}{104053} (\bibinfo{year}{2012}), \eprint{1208.3473}.

\bibitem[{\citenamefont{Moldenhauer et~al.}(2014)\citenamefont{Moldenhauer,
  Markakis, Johnson-McDaniel, Tichy, and Brügmann}}]{Moldenhauer:2014yaa}
\bibinfo{author}{\bibfnamefont{N.}~\bibnamefont{Moldenhauer}},
  \bibinfo{author}{\bibfnamefont{C.~M.} \bibnamefont{Markakis}},
  \bibinfo{author}{\bibfnamefont{N.~K.} \bibnamefont{Johnson-McDaniel}},
  \bibinfo{author}{\bibfnamefont{W.}~\bibnamefont{Tichy}}, \bibnamefont{and}
  \bibinfo{author}{\bibfnamefont{B.}~\bibnamefont{Brügmann}},
  \bibinfo{journal}{Phys. Rev.} \textbf{\bibinfo{volume}{D90}},
  \bibinfo{pages}{084043} (\bibinfo{year}{2014}), \eprint{1408.4136}.

\bibitem[{\citenamefont{Dietrich
  et~al.}(2019{\natexlab{a}})\citenamefont{Dietrich, Ossokine, and
  Clough}}]{Dietrich:2018bvi}
\bibinfo{author}{\bibfnamefont{T.}~\bibnamefont{Dietrich}},
  \bibinfo{author}{\bibfnamefont{S.}~\bibnamefont{Ossokine}}, \bibnamefont{and}
  \bibinfo{author}{\bibfnamefont{K.}~\bibnamefont{Clough}},
  \bibinfo{journal}{Class. Quant. Grav.} \textbf{\bibinfo{volume}{36}},
  \bibinfo{pages}{025002} (\bibinfo{year}{2019}{\natexlab{a}}),
  \eprint{1807.06959}.

\bibitem[{\citenamefont{Tsokaros and Uryu}(2012)}]{Tsokaros:2012kp}
\bibinfo{author}{\bibfnamefont{A.}~\bibnamefont{Tsokaros}} \bibnamefont{and}
  \bibinfo{author}{\bibfnamefont{K.}~\bibnamefont{Uryu}}, \bibinfo{journal}{J.
  Eng. Math.} \textbf{\bibinfo{volume}{82}}, \bibinfo{pages}{1}
  (\bibinfo{year}{2012}), \eprint{1207.5833}.

\bibitem[{\citenamefont{Tsokaros et~al.}(2015)\citenamefont{Tsokaros, Uryu, and
  Rezzolla}}]{Tsokaros:2015fea}
\bibinfo{author}{\bibfnamefont{A.}~\bibnamefont{Tsokaros}},
  \bibinfo{author}{\bibfnamefont{K.}~\bibnamefont{Uryu}}, \bibnamefont{and}
  \bibinfo{author}{\bibfnamefont{L.}~\bibnamefont{Rezzolla}},
  \bibinfo{journal}{Phys. Rev.} \textbf{\bibinfo{volume}{D91}},
  \bibinfo{pages}{104030} (\bibinfo{year}{2015}), \eprint{1502.05674}.

\bibitem[{\citenamefont{Foucart et~al.}(2008)\citenamefont{Foucart, Kidder,
  Pfeiffer, and Teukolsky}}]{Foucart:2008qt}
\bibinfo{author}{\bibfnamefont{F.}~\bibnamefont{Foucart}},
  \bibinfo{author}{\bibfnamefont{L.~E.} \bibnamefont{Kidder}},
  \bibinfo{author}{\bibfnamefont{H.~P.} \bibnamefont{Pfeiffer}},
  \bibnamefont{and} \bibinfo{author}{\bibfnamefont{S.~A.}
  \bibnamefont{Teukolsky}}, \bibinfo{journal}{Phys. Rev.}
  \textbf{\bibinfo{volume}{D77}}, \bibinfo{pages}{124051}
  (\bibinfo{year}{2008}), \eprint{0804.3787}.

\bibitem[{\citenamefont{Tacik et~al.}(2015)}]{Tacik:2015tja}
\bibinfo{author}{\bibfnamefont{N.}~\bibnamefont{Tacik}} \bibnamefont{et~al.},
  \bibinfo{journal}{Phys. Rev.} \textbf{\bibinfo{volume}{D92}},
  \bibinfo{pages}{124012} (\bibinfo{year}{2015}), \eprint{1508.06986}.

\bibitem[{\citenamefont{Tichy}(2006)}]{Tichy:2006qn}
\bibinfo{author}{\bibfnamefont{W.}~\bibnamefont{Tichy}},
  \bibinfo{journal}{Phys. Rev.} \textbf{\bibinfo{volume}{D74}},
  \bibinfo{pages}{084005} (\bibinfo{year}{2006}), \eprint{gr-qc/0609087}.

\bibitem[{\citenamefont{Tichy}(2009{\natexlab{a}})}]{Tichy:2009yr}
\bibinfo{author}{\bibfnamefont{W.}~\bibnamefont{Tichy}},
  \bibinfo{journal}{Class. Quant. Grav.} \textbf{\bibinfo{volume}{26}},
  \bibinfo{pages}{175018} (\bibinfo{year}{2009}{\natexlab{a}}),
  \eprint{0908.0620}.

\bibitem[{\citenamefont{Tichy}(2009{\natexlab{b}})}]{Tichy:2009zr}
\bibinfo{author}{\bibfnamefont{W.}~\bibnamefont{Tichy}},
  \bibinfo{journal}{Phys. Rev.} \textbf{\bibinfo{volume}{D80}},
  \bibinfo{pages}{104034} (\bibinfo{year}{2009}{\natexlab{b}}),
  \eprint{0911.0973}.

\bibitem[{\citenamefont{Rüter et~al.}(2018)\citenamefont{Rüter, Hilditch,
  Bugner, and Brügmann}}]{Ruter:2017iph}
\bibinfo{author}{\bibfnamefont{H.~R.} \bibnamefont{Rüter}},
  \bibinfo{author}{\bibfnamefont{D.}~\bibnamefont{Hilditch}},
  \bibinfo{author}{\bibfnamefont{M.}~\bibnamefont{Bugner}}, \bibnamefont{and}
  \bibinfo{author}{\bibfnamefont{B.}~\bibnamefont{Brügmann}},
  \bibinfo{journal}{Phys. Rev.} \textbf{\bibinfo{volume}{D98}},
  \bibinfo{pages}{084044} (\bibinfo{year}{2018}), \eprint{1708.07358}.

\bibitem[{\citenamefont{Vincent et~al.}(2019)\citenamefont{Vincent, Pfeiffer,
  and Fischer}}]{Vincent:2019qpd}
\bibinfo{author}{\bibfnamefont{T.}~\bibnamefont{Vincent}},
  \bibinfo{author}{\bibfnamefont{H.~P.} \bibnamefont{Pfeiffer}},
  \bibnamefont{and} \bibinfo{author}{\bibfnamefont{N.~L.}
  \bibnamefont{Fischer}} (\bibinfo{year}{2019}), \eprint{1907.01572}.

\bibitem[{\citenamefont{Bernuzzi et~al.}(2014)\citenamefont{Bernuzzi, Dietrich,
  Tichy, and Brügmann}}]{Bernuzzi:2013rza}
\bibinfo{author}{\bibfnamefont{S.}~\bibnamefont{Bernuzzi}},
  \bibinfo{author}{\bibfnamefont{T.}~\bibnamefont{Dietrich}},
  \bibinfo{author}{\bibfnamefont{W.}~\bibnamefont{Tichy}}, \bibnamefont{and}
  \bibinfo{author}{\bibfnamefont{B.}~\bibnamefont{Brügmann}},
  \bibinfo{journal}{Phys.Rev.} \textbf{\bibinfo{volume}{D89}},
  \bibinfo{pages}{104021} (\bibinfo{year}{2014}), \eprint{1311.4443}.

\bibitem[{\citenamefont{Dietrich
  et~al.}(2018{\natexlab{a}})\citenamefont{Dietrich, Bernuzzi, Brügmann, and
  Tichy}}]{Dietrich:2018upm}
\bibinfo{author}{\bibfnamefont{T.}~\bibnamefont{Dietrich}},
  \bibinfo{author}{\bibfnamefont{S.}~\bibnamefont{Bernuzzi}},
  \bibinfo{author}{\bibfnamefont{B.}~\bibnamefont{Brügmann}},
  \bibnamefont{and} \bibinfo{author}{\bibfnamefont{W.}~\bibnamefont{Tichy}}, in
  \emph{\bibinfo{booktitle}{{Proceedings, 26th Euromicro International
  Conference on Parallel, Distributed and Network-based Processing (PDP 2018):
  Cambridge, UK, March 21-23, 2018}}} (\bibinfo{year}{2018}{\natexlab{a}}), pp.
  \bibinfo{pages}{682--689}, \eprint{1803.07965}.

\bibitem[{\citenamefont{Most et~al.}(2019)\citenamefont{Most, Papenfort,
  Tsokaros, and Rezzolla}}]{Most:2019pac}
\bibinfo{author}{\bibfnamefont{E.~R.} \bibnamefont{Most}},
  \bibinfo{author}{\bibfnamefont{L.~J.} \bibnamefont{Papenfort}},
  \bibinfo{author}{\bibfnamefont{A.}~\bibnamefont{Tsokaros}}, \bibnamefont{and}
  \bibinfo{author}{\bibfnamefont{L.}~\bibnamefont{Rezzolla}}
  (\bibinfo{year}{2019}), \eprint{1904.04220}.

\bibitem[{\citenamefont{Tsokaros et~al.}(2019)\citenamefont{Tsokaros, Ruiz,
  Paschalidis, Shapiro, and Uryu}}]{Tsokaros:2019anx}
\bibinfo{author}{\bibfnamefont{A.}~\bibnamefont{Tsokaros}},
  \bibinfo{author}{\bibfnamefont{M.}~\bibnamefont{Ruiz}},
  \bibinfo{author}{\bibfnamefont{V.}~\bibnamefont{Paschalidis}},
  \bibinfo{author}{\bibfnamefont{S.~L.} \bibnamefont{Shapiro}},
  \bibnamefont{and} \bibinfo{author}{\bibfnamefont{K.}~\bibnamefont{Uryu}}
  (\bibinfo{year}{2019}), \eprint{1906.00011}.

\bibitem[{\citenamefont{East et~al.}(2019)\citenamefont{East, Paschalidis,
  Pretorius, and Tsokaros}}]{East:2019lbk}
\bibinfo{author}{\bibfnamefont{W.~E.} \bibnamefont{East}},
  \bibinfo{author}{\bibfnamefont{V.}~\bibnamefont{Paschalidis}},
  \bibinfo{author}{\bibfnamefont{F.}~\bibnamefont{Pretorius}},
  \bibnamefont{and} \bibinfo{author}{\bibfnamefont{A.}~\bibnamefont{Tsokaros}}
  (\bibinfo{year}{2019}), \eprint{1906.05288}.

\bibitem[{\citenamefont{Dietrich
  et~al.}(2018{\natexlab{b}})\citenamefont{Dietrich, Bernuzzi, Brügmann,
  Ujevic, and Tichy}}]{Dietrich:2017xqb}
\bibinfo{author}{\bibfnamefont{T.}~\bibnamefont{Dietrich}},
  \bibinfo{author}{\bibfnamefont{S.}~\bibnamefont{Bernuzzi}},
  \bibinfo{author}{\bibfnamefont{B.}~\bibnamefont{Brügmann}},
  \bibinfo{author}{\bibfnamefont{M.}~\bibnamefont{Ujevic}}, \bibnamefont{and}
  \bibinfo{author}{\bibfnamefont{W.}~\bibnamefont{Tichy}},
  \bibinfo{journal}{Phys. Rev.} \textbf{\bibinfo{volume}{D97}},
  \bibinfo{pages}{064002} (\bibinfo{year}{2018}{\natexlab{b}}),
  \eprint{1712.02992}.

\bibitem[{\citenamefont{Foucart et~al.}(2016)\citenamefont{Foucart, Haas, Duez,
  O'Connor, Ott, Roberts, Kidder, Lippuner, Pfeiffer, and
  Scheel}}]{Foucart:2015gaa}
\bibinfo{author}{\bibfnamefont{F.}~\bibnamefont{Foucart}},
  \bibinfo{author}{\bibfnamefont{R.}~\bibnamefont{Haas}},
  \bibinfo{author}{\bibfnamefont{M.~D.} \bibnamefont{Duez}},
  \bibinfo{author}{\bibfnamefont{E.}~\bibnamefont{O'Connor}},
  \bibinfo{author}{\bibfnamefont{C.~D.} \bibnamefont{Ott}},
  \bibinfo{author}{\bibfnamefont{L.}~\bibnamefont{Roberts}},
  \bibinfo{author}{\bibfnamefont{L.~E.} \bibnamefont{Kidder}},
  \bibinfo{author}{\bibfnamefont{J.}~\bibnamefont{Lippuner}},
  \bibinfo{author}{\bibfnamefont{H.~P.} \bibnamefont{Pfeiffer}},
  \bibnamefont{and} \bibinfo{author}{\bibfnamefont{M.~A.}
  \bibnamefont{Scheel}}, \bibinfo{journal}{Phys. Rev.}
  \textbf{\bibinfo{volume}{D93}}, \bibinfo{pages}{044019}
  (\bibinfo{year}{2016}), \eprint{1510.06398}.

\bibitem[{\citenamefont{Kiuchi et~al.}(2017)\citenamefont{Kiuchi, Kawaguchi,
  Kyutoku, Sekiguchi, Shibata, and Taniguchi}}]{Kiuchi:2017pte}
\bibinfo{author}{\bibfnamefont{K.}~\bibnamefont{Kiuchi}},
  \bibinfo{author}{\bibfnamefont{K.}~\bibnamefont{Kawaguchi}},
  \bibinfo{author}{\bibfnamefont{K.}~\bibnamefont{Kyutoku}},
  \bibinfo{author}{\bibfnamefont{Y.}~\bibnamefont{Sekiguchi}},
  \bibinfo{author}{\bibfnamefont{M.}~\bibnamefont{Shibata}}, \bibnamefont{and}
  \bibinfo{author}{\bibfnamefont{K.}~\bibnamefont{Taniguchi}},
  \bibinfo{journal}{Phys. Rev.} \textbf{\bibinfo{volume}{D96}},
  \bibinfo{pages}{084060} (\bibinfo{year}{2017}), \eprint{1708.08926}.

\bibitem[{\citenamefont{Dietrich
  et~al.}(2017{\natexlab{b}})\citenamefont{Dietrich, Bernuzzi, and
  Tichy}}]{Dietrich:2017aum}
\bibinfo{author}{\bibfnamefont{T.}~\bibnamefont{Dietrich}},
  \bibinfo{author}{\bibfnamefont{S.}~\bibnamefont{Bernuzzi}}, \bibnamefont{and}
  \bibinfo{author}{\bibfnamefont{W.}~\bibnamefont{Tichy}},
  \bibinfo{journal}{Phys. Rev.} \textbf{\bibinfo{volume}{D96}},
  \bibinfo{pages}{121501} (\bibinfo{year}{2017}{\natexlab{b}}),
  \eprint{1706.02969}.

\bibitem[{\citenamefont{Foucart et~al.}(2019)}]{Foucart:2018lhe}
\bibinfo{author}{\bibfnamefont{F.}~\bibnamefont{Foucart}} \bibnamefont{et~al.},
  \bibinfo{journal}{Phys. Rev.} \textbf{\bibinfo{volume}{D99}},
  \bibinfo{pages}{044008} (\bibinfo{year}{2019}), \eprint{1812.06988}.

\bibitem[{\citenamefont{Dietrich
  et~al.}(2019{\natexlab{b}})\citenamefont{Dietrich, Samajdar, Khan,
  Johnson-McDaniel, Dudi, and Tichy}}]{Dietrich:2019kaq}
\bibinfo{author}{\bibfnamefont{T.}~\bibnamefont{Dietrich}},
  \bibinfo{author}{\bibfnamefont{A.}~\bibnamefont{Samajdar}},
  \bibinfo{author}{\bibfnamefont{S.}~\bibnamefont{Khan}},
  \bibinfo{author}{\bibfnamefont{N.~K.} \bibnamefont{Johnson-McDaniel}},
  \bibinfo{author}{\bibfnamefont{R.}~\bibnamefont{Dudi}}, \bibnamefont{and}
  \bibinfo{author}{\bibfnamefont{W.}~\bibnamefont{Tichy}}
  (\bibinfo{year}{2019}{\natexlab{b}}), \eprint{1905.06011}.

\bibitem[{\citenamefont{Kiuchi et~al.}(2019{\natexlab{b}})\citenamefont{Kiuchi,
  Kyohei, Kyutoku, Sekiguchi, and Shibata}}]{Kiuchi:2019kzt}
\bibinfo{author}{\bibfnamefont{K.}~\bibnamefont{Kiuchi}},
  \bibinfo{author}{\bibfnamefont{K.}~\bibnamefont{Kyohei}},
  \bibinfo{author}{\bibfnamefont{K.}~\bibnamefont{Kyutoku}},
  \bibinfo{author}{\bibfnamefont{Y.}~\bibnamefont{Sekiguchi}},
  \bibnamefont{and} \bibinfo{author}{\bibfnamefont{M.}~\bibnamefont{Shibata}}
  (\bibinfo{year}{2019}{\natexlab{b}}), \eprint{1907.03790}.

\bibitem[{\citenamefont{Chaurasia et~al.}(2018)\citenamefont{Chaurasia,
  Dietrich, Johnson-McDaniel, Ujevic, Tichy, and
  Brügmann}}]{Chaurasia:2018zhg}
\bibinfo{author}{\bibfnamefont{S.~V.} \bibnamefont{Chaurasia}},
  \bibinfo{author}{\bibfnamefont{T.}~\bibnamefont{Dietrich}},
  \bibinfo{author}{\bibfnamefont{N.~K.} \bibnamefont{Johnson-McDaniel}},
  \bibinfo{author}{\bibfnamefont{M.}~\bibnamefont{Ujevic}},
  \bibinfo{author}{\bibfnamefont{W.}~\bibnamefont{Tichy}}, \bibnamefont{and}
  \bibinfo{author}{\bibfnamefont{B.}~\bibnamefont{Brügmann}},
  \bibinfo{journal}{Phys. Rev.} \textbf{\bibinfo{volume}{D98}},
  \bibinfo{pages}{104005} (\bibinfo{year}{2018}), \eprint{1807.06857}.

\bibitem[{\citenamefont{Dietrich
  et~al.}(2018{\natexlab{c}})\citenamefont{Dietrich, Radice, Bernuzzi, Zappa,
  Perego, Brügmann, Chaurasia, Dudi, Tichy, and Ujevic}}]{Dietrich:2018phi}
\bibinfo{author}{\bibfnamefont{T.}~\bibnamefont{Dietrich}},
  \bibinfo{author}{\bibfnamefont{D.}~\bibnamefont{Radice}},
  \bibinfo{author}{\bibfnamefont{S.}~\bibnamefont{Bernuzzi}},
  \bibinfo{author}{\bibfnamefont{F.}~\bibnamefont{Zappa}},
  \bibinfo{author}{\bibfnamefont{A.}~\bibnamefont{Perego}},
  \bibinfo{author}{\bibfnamefont{B.}~\bibnamefont{Brügmann}},
  \bibinfo{author}{\bibfnamefont{S.~V.} \bibnamefont{Chaurasia}},
  \bibinfo{author}{\bibfnamefont{R.}~\bibnamefont{Dudi}},
  \bibinfo{author}{\bibfnamefont{W.}~\bibnamefont{Tichy}}, \bibnamefont{and}
  \bibinfo{author}{\bibfnamefont{M.}~\bibnamefont{Ujevic}},
  \bibinfo{journal}{Class. Quant. Grav.} \textbf{\bibinfo{volume}{35}},
  \bibinfo{pages}{24LT01} (\bibinfo{year}{2018}{\natexlab{c}}),
  \eprint{1806.01625}.

\bibitem[{\citenamefont{Radice et~al.}(2018)\citenamefont{Radice, Perego,
  Hotokezaka, Fromm, Bernuzzi, and Roberts}}]{Radice:2018pdn}
\bibinfo{author}{\bibfnamefont{D.}~\bibnamefont{Radice}},
  \bibinfo{author}{\bibfnamefont{A.}~\bibnamefont{Perego}},
  \bibinfo{author}{\bibfnamefont{K.}~\bibnamefont{Hotokezaka}},
  \bibinfo{author}{\bibfnamefont{S.~A.} \bibnamefont{Fromm}},
  \bibinfo{author}{\bibfnamefont{S.}~\bibnamefont{Bernuzzi}}, \bibnamefont{and}
  \bibinfo{author}{\bibfnamefont{L.~F.} \bibnamefont{Roberts}},
  \bibinfo{journal}{Astrophys. J.} \textbf{\bibinfo{volume}{869}},
  \bibinfo{pages}{130} (\bibinfo{year}{2018}), \eprint{1809.11161}.

\bibitem[{\citenamefont{Dietrich
  et~al.}(2017{\natexlab{c}})\citenamefont{Dietrich, Ujevic, Tichy, Bernuzzi,
  and Brügmann}}]{Dietrich:2016hky}
\bibinfo{author}{\bibfnamefont{T.}~\bibnamefont{Dietrich}},
  \bibinfo{author}{\bibfnamefont{M.}~\bibnamefont{Ujevic}},
  \bibinfo{author}{\bibfnamefont{W.}~\bibnamefont{Tichy}},
  \bibinfo{author}{\bibfnamefont{S.}~\bibnamefont{Bernuzzi}}, \bibnamefont{and}
  \bibinfo{author}{\bibfnamefont{B.}~\bibnamefont{Brügmann}},
  \bibinfo{journal}{Phys. Rev.} \textbf{\bibinfo{volume}{D95}},
  \bibinfo{pages}{024029} (\bibinfo{year}{2017}{\natexlab{c}}),
  \eprint{1607.06636}.

\bibitem[{\citenamefont{Abbott et~al.}(2019{\natexlab{a}})}]{Abbott:2018wiz}
\bibinfo{author}{\bibfnamefont{B.~P.} \bibnamefont{Abbott}}
  \bibnamefont{et~al.} (\bibinfo{collaboration}{LIGO Scientific, Virgo}),
  \bibinfo{journal}{Phys. Rev.} \textbf{\bibinfo{volume}{X9}},
  \bibinfo{pages}{011001} (\bibinfo{year}{2019}{\natexlab{a}}),
  \eprint{1805.11579}.

\bibitem[{\citenamefont{Abbott et~al.}(2018)}]{Abbott:2018exr}
\bibinfo{author}{\bibfnamefont{B.~P.} \bibnamefont{Abbott}}
  \bibnamefont{et~al.} (\bibinfo{collaboration}{Virgo, LIGO Scientific}),
  \bibinfo{journal}{Phys. Rev. Lett.} \textbf{\bibinfo{volume}{121}},
  \bibinfo{pages}{161101} (\bibinfo{year}{2018}), \eprint{1805.11581}.

\bibitem[{\citenamefont{Abbott et~al.}(2019{\natexlab{b}})}]{Abbott:2018lct}
\bibinfo{author}{\bibfnamefont{B.~P.} \bibnamefont{Abbott}}
  \bibnamefont{et~al.} (\bibinfo{collaboration}{LIGO Scientific, Virgo}),
  \bibinfo{journal}{Phys. Rev. Lett.} \textbf{\bibinfo{volume}{123}},
  \bibinfo{pages}{011102} (\bibinfo{year}{2019}{\natexlab{b}}),
  \eprint{1811.00364}.

\bibitem[{\citenamefont{Tichy}(2011)}]{Tichy:2011gw}
\bibinfo{author}{\bibfnamefont{W.}~\bibnamefont{Tichy}},
  \bibinfo{journal}{Phys. Rev.} \textbf{\bibinfo{volume}{D84}},
  \bibinfo{pages}{024041} (\bibinfo{year}{2011}), \eprint{1107.1440}.

\bibitem[{\citenamefont{Tichy}(2012)}]{Tichy:2012rp}
\bibinfo{author}{\bibfnamefont{W.}~\bibnamefont{Tichy}},
  \bibinfo{journal}{Phys.Rev.} \textbf{\bibinfo{volume}{D86}},
  \bibinfo{pages}{064024} (\bibinfo{year}{2012}), \eprint{1209.5336}.

\bibitem[{\citenamefont{Rüter}(2019)}]{rueter_2019}
\bibinfo{author}{\bibfnamefont{H.}~\bibnamefont{Rüter}}, Ph.D. thesis,
  \bibinfo{school}{Friedrich-Schiller-Universität Jena}
  (\bibinfo{year}{2019}),
  \urlprefix\url{https://www.db-thueringen.de/receive/dbt_mods_00039088}.

\bibitem[{\citenamefont{Arnowitt et~al.}(1962)\citenamefont{Arnowitt, Deser,
  and Misner}}]{Arnowitt62}
\bibinfo{author}{\bibfnamefont{R.}~\bibnamefont{Arnowitt}},
  \bibinfo{author}{\bibfnamefont{S.}~\bibnamefont{Deser}}, \bibnamefont{and}
  \bibinfo{author}{\bibfnamefont{C.~W.} \bibnamefont{Misner}}, in
  \emph{\bibinfo{booktitle}{Gravitation: An Introduction to Current Research}},
  edited by \bibinfo{editor}{\bibfnamefont{L.}~\bibnamefont{Witten}}
  (\bibinfo{publisher}{John Wiley}, \bibinfo{address}{New York},
  \bibinfo{year}{1962}), pp. \bibinfo{pages}{227--265},
  \bibinfo{note}{arXiv:gr-qc/0405109}, \eprint{gr-qc/0405109}.

\bibitem[{\citenamefont{Read et~al.}(2009)\citenamefont{Read, Lackey, Owen, and
  Friedman}}]{Read:2008iy}
\bibinfo{author}{\bibfnamefont{J.~S.} \bibnamefont{Read}},
  \bibinfo{author}{\bibfnamefont{B.~D.} \bibnamefont{Lackey}},
  \bibinfo{author}{\bibfnamefont{B.~J.} \bibnamefont{Owen}}, \bibnamefont{and}
  \bibinfo{author}{\bibfnamefont{J.~L.} \bibnamefont{Friedman}},
  \bibinfo{journal}{Phys.Rev.} \textbf{\bibinfo{volume}{D79}},
  \bibinfo{pages}{124032} (\bibinfo{year}{2009}), \eprint{0812.2163}.

\bibitem[{\citenamefont{Ansorg}(2007)}]{Ansorg:2006gd}
\bibinfo{author}{\bibfnamefont{M.}~\bibnamefont{Ansorg}},
  \bibinfo{journal}{Class. Quant. Grav.} \textbf{\bibinfo{volume}{24}},
  \bibinfo{pages}{S1} (\bibinfo{year}{2007}), \eprint{gr-qc/0612081}.

\bibitem[{\citenamefont{Reifenberger}(2013)}]{Reifenberger2013}
\bibinfo{author}{\bibfnamefont{G.}~\bibnamefont{Reifenberger}}, Ph.D. thesis,
  \bibinfo{school}{Florida Atlantic University} (\bibinfo{year}{2013}).

\bibitem[{\citenamefont{Davis and Duff}(1997)}]{Davis-Duff-1997-UMFPACK}
\bibinfo{author}{\bibfnamefont{T.~A.} \bibnamefont{Davis}} \bibnamefont{and}
  \bibinfo{author}{\bibfnamefont{I.~S.} \bibnamefont{Duff}},
  \bibinfo{journal}{SIAM J. Matrix Anal. Applic.}
  \textbf{\bibinfo{volume}{18}}, \bibinfo{pages}{140} (\bibinfo{year}{1997}).

\bibitem[{\citenamefont{Davis and Duff}(1999)}]{Davis-Duff_UMFPACK_1999}
\bibinfo{author}{\bibfnamefont{T.~A.} \bibnamefont{Davis}} \bibnamefont{and}
  \bibinfo{author}{\bibfnamefont{I.~S.} \bibnamefont{Duff}},
  \bibinfo{journal}{ACM Trans. Math. Softw.} \textbf{\bibinfo{volume}{25}},
  \bibinfo{pages}{1} (\bibinfo{year}{1999}), ISSN \bibinfo{issn}{0098-3500}.

\bibitem[{\citenamefont{Davis}(2004{\natexlab{a}})}]{Davis_UMFPACK_V4.3_2004}
\bibinfo{author}{\bibfnamefont{T.~A.} \bibnamefont{Davis}},
  \bibinfo{journal}{ACM Trans. Math. Softw.} \textbf{\bibinfo{volume}{30}},
  \bibinfo{pages}{196} (\bibinfo{year}{2004}{\natexlab{a}}), ISSN
  \bibinfo{issn}{0098-3500}.

\bibitem[{\citenamefont{Davis}(2004{\natexlab{b}})}]{Davis_UMFPACK_2004}
\bibinfo{author}{\bibfnamefont{T.~A.} \bibnamefont{Davis}},
  \bibinfo{journal}{ACM Trans. Math. Softw.} \textbf{\bibinfo{volume}{30}},
  \bibinfo{pages}{165} (\bibinfo{year}{2004}{\natexlab{b}}), ISSN
  \bibinfo{issn}{0098-3500}.

\bibitem[{\citenamefont{Davis}()}]{umfpack_web}
\bibinfo{author}{\bibfnamefont{T.~A.} \bibnamefont{Davis}},
  \bibinfo{note}{{UMFPACK} a sparse linear systems solver using the Unsymmetric
  MultiFrontal method:\\ {\tt
  http://www.cise.ufl.edu/research/sparse/umfpack/}}.

\bibitem[{\citenamefont{Gourgoulhon}(2007)}]{Gourgoulhon:2007ue}
\bibinfo{author}{\bibfnamefont{E.}~\bibnamefont{Gourgoulhon}}
  (\bibinfo{year}{2007}), \eprint{gr-qc/0703035}.

\bibitem[{\citenamefont{Misner et~al.}(1973)\citenamefont{Misner, Thorne, and
  Wheeler}}]{Misner73}
\bibinfo{author}{\bibfnamefont{C.~W.} \bibnamefont{Misner}},
  \bibinfo{author}{\bibfnamefont{K.~S.} \bibnamefont{Thorne}},
  \bibnamefont{and} \bibinfo{author}{\bibfnamefont{J.~A.}
  \bibnamefont{Wheeler}}, \emph{\bibinfo{title}{Gravitation}}
  (\bibinfo{publisher}{W. H. Freeman}, \bibinfo{address}{San Francisco},
  \bibinfo{year}{1973}).

\bibitem[{\citenamefont{Campanelli et~al.}(2007)\citenamefont{Campanelli,
  Lousto, Zlochower, Krishnan, and Merritt}}]{Campanelli:2006fy}
\bibinfo{author}{\bibfnamefont{M.}~\bibnamefont{Campanelli}},
  \bibinfo{author}{\bibfnamefont{C.~O.} \bibnamefont{Lousto}},
  \bibinfo{author}{\bibfnamefont{Y.}~\bibnamefont{Zlochower}},
  \bibinfo{author}{\bibfnamefont{B.}~\bibnamefont{Krishnan}}, \bibnamefont{and}
  \bibinfo{author}{\bibfnamefont{D.}~\bibnamefont{Merritt}},
  \bibinfo{journal}{Phys. Rev.} \textbf{\bibinfo{volume}{D75}},
  \bibinfo{pages}{064030} (\bibinfo{year}{2007}), \eprint{gr-qc/0612076}.

\bibitem[{\citenamefont{Ansorg et~al.}(2003)\citenamefont{Ansorg,
  Kleinw\"achter, and Meinel}}]{Ansorg:2003br}
\bibinfo{author}{\bibfnamefont{M.}~\bibnamefont{Ansorg}},
  \bibinfo{author}{\bibfnamefont{A.}~\bibnamefont{Kleinw\"achter}},
  \bibnamefont{and} \bibinfo{author}{\bibfnamefont{R.}~\bibnamefont{Meinel}},
  \bibinfo{journal}{Astron. Astrophys.} \textbf{\bibinfo{volume}{405}},
  \bibinfo{pages}{711} (\bibinfo{year}{2003}), \eprint{astro-ph/0301173}.

\bibitem[{\citenamefont{Damour et~al.}(2012)\citenamefont{Damour, Nagar,
  Pollney, and Reisswig}}]{Damour:2011fu}
\bibinfo{author}{\bibfnamefont{T.}~\bibnamefont{Damour}},
  \bibinfo{author}{\bibfnamefont{A.}~\bibnamefont{Nagar}},
  \bibinfo{author}{\bibfnamefont{D.}~\bibnamefont{Pollney}}, \bibnamefont{and}
  \bibinfo{author}{\bibfnamefont{C.}~\bibnamefont{Reisswig}},
  \bibinfo{journal}{Phys. Rev. Lett.} \textbf{\bibinfo{volume}{108}},
  \bibinfo{pages}{131101} (\bibinfo{year}{2012}), \eprint{1110.2938}.

\bibitem[{\citenamefont{Dietrich and Hinderer}(2017)}]{Dietrich:2017feu}
\bibinfo{author}{\bibfnamefont{T.}~\bibnamefont{Dietrich}} \bibnamefont{and}
  \bibinfo{author}{\bibfnamefont{T.}~\bibnamefont{Hinderer}},
  \bibinfo{journal}{Phys. Rev.} \textbf{\bibinfo{volume}{D95}},
  \bibinfo{pages}{124006} (\bibinfo{year}{2017}), \eprint{1702.02053}.

\bibitem[{\citenamefont{Marronetti and Shapiro}(2003)}]{Marronetti:2003gk}
\bibinfo{author}{\bibfnamefont{P.}~\bibnamefont{Marronetti}} \bibnamefont{and}
  \bibinfo{author}{\bibfnamefont{S.~L.} \bibnamefont{Shapiro}},
  \bibinfo{journal}{Phys. Rev.} \textbf{\bibinfo{volume}{D68}},
  \bibinfo{pages}{104024} (\bibinfo{year}{2003}), \eprint{gr-qc/0306075}.

\end{thebibliography}

\end{document}